\documentclass[final,3p,authoryear,12pt]{elsarticle}

\usepackage[page]{appendix}

\fontfamily{cmss}\selectfont
\usepackage{inconsolata}
\usepackage{setspace} 
\setcitestyle{square,numbers}
\usepackage[colorinlistoftodos,textsize=footnotesize]{todonotes} 

\usepackage{amsfonts}
\usepackage{amsmath}
\usepackage{amssymb}
\usepackage{amsthm}
\usepackage{lscape} 
\usepackage{float}
\usepackage{graphicx}
\usepackage{tabularx}
\usepackage{url}
\usepackage{times}
\usepackage{color}
\usepackage{afterpage}
\usepackage{rotating}
\usepackage{hyperref}
\usepackage{xcolor}
\usepackage{tikz}
\usepackage{moresize}
\usepackage{multirow}
\usepackage{array}
\usepackage{enumerate}
\usepackage{colortbl}
\usepackage{bm}
\usepackage{bbm}
\usepackage{diagbox}
\usepackage{booktabs}
\usepackage{floatpag}
\usepackage{accents}
\usepackage[utf8]{inputenc}
\usepackage{mathtools}
\usepackage{xfrac}
\usepackage[caption=true]{subfig}
\usepackage[tableposition=top, skip=10pt]{caption}
\usepackage[ruled, vlined]{algorithm2e}
\usepackage[inline]{enumitem}
\usepackage{soul}
\usepackage[normalem]{ulem}
\usepackage{nicefrac} 
\usepackage{doi}
\usepackage{diagbox}
\usepackage[bottom]{footmisc}
\usepackage{cleveref}

\usetikzlibrary{calc}

    \SetKwComment{Comment}{$\triangleright$\ }{}
    \makeatletter
        \def\algbackskip{\hskip-\ALG@thistlm}
    \makeatother
    
    \SetCommentSty{mycommfont}
    
\allowdisplaybreaks

\newcommand{\N}{\mathbb{N}}
\newcommand{\R}{\mathbb{R}}
\newcommand{\E}{\mathbb{E}}
\newcommand{\bbP}{\mathbb{P}}
\newcommand{\bbbP}{\bar{\mathbb{P}}}
\newcommand{\bbQ}{\mathbb{Q}}

\newcommand{\bfC}{\mathbf{C}}
\newcommand{\bfX}{\mathbf{X}}
\newcommand{\bfY}{\mathbf{Y}}
\newcommand{\bfS}{\mathbf{S}}
\newcommand{\bfW}{\mathbf{W}}

\newcommand{\bz}{\bar{z}}
\newcommand{\bphi}{\bar{\phi}}

\newcommand{\calF}{\mathcal{F}}

\newcommand{\calP}{\mathcal{P}}
\newcommand{\calG}{\mathcal{G}}
\newcommand{\calL}{\mathcal{L}}

\newcommand{\calC}{\mathcal{C}}

\newcommand{\calX}{\mathcal{X}}

\newcommand{\frakF}{\mathfrak{F}}
\newcommand{\PnL}{\mathrm{P\&L}}

\newcommand{\myRM}{U}

\DeclareMathOperator*{\argmin}{\mathrm{arg\,min}}

\newcommand{\TI}{J}
\newcommand{\sig}{\mathrm{Sig}}
\newcommand{\sigw}{\mathrm{SigW}_1}
\newcommand{\sigmmd}{\mathrm{SigMMD}}
\newcommand{\dist}{\mathrm{dist}}
\newcommand{\bdot}{\boldsymbol{\cdot}}

\newcommand{\myE}[2]{\E_{#1}\left[ #2 \right]}
\newcommand{\myV}[2]{\mathbb{V}_{#1}\left[ #2 \right]}
\newcommand{\myCond}[3]{\E_{#1}\left[\left. #2 \right\vert #3 \right]}

\newcommand{\mySet}[2]{\left\lbrace #1 : #2 \right\rbrace}
\newcommand{\myset}[2]{{\lbrace #1, \ldots, #2 \rbrace}}

\definecolor{myblue}{rgb}{0.8,0.8,1}
\definecolor{myred}{rgb}{1,0.8,0.8}
\definecolor{mygreen}{rgb}{0.8,1,0.8}
\definecolor{mygrey}{rgb}{220,220,220}

\definecolor{dbblue}{RGB}{10,65,155}				
\definecolor{dbred}{RGB}{215,0,50}					
\definecolor{blue}{RGB}{0,113.9850,188.9550} 			
\definecolor{red}{RGB}{216.7500,82.8750,24.9900} 		
\definecolor{green}{RGB}{118.8300,171.8700,47.9400} 	
\definecolor{grey}{RGB}{110,110,110}				
\definecolor{lgrey}{RGB}{210,210,210}				
\definecolor{c1}{RGB}{0,113.9850,188.9550}			
\definecolor{c2}{RGB}{216.7500,82.8750,24.9900}		
\definecolor{c3}{RGB}{236.8950,176.9700,31.8750}		
\definecolor{c4}{RGB}{125.9700,46.9200,141.7800}		
\definecolor{c5}{RGB}{118.8300,171.8700,47.9400}		
\definecolor{c6}{RGB}{76.7550,189.9750,237.9150}		
\definecolor{c7}{RGB}{161.9250,19.8900,46.9200}		
\definecolor{c14}{RGB}{0,102,102}					

\definecolor{C0}{rgb}{0.2980392156862745, 0.4470588235294118, 0.6901960784313725}
\definecolor{C1}{rgb}{0.8666666666666667, 0.5176470588235295, 0.3215686274509804}
\definecolor{C2}{rgb}{0.3333333333333333, 0.6588235294117647, 0.4078431372549019}
\definecolor{C3}{rgb}{0.7686274509803922, 0.3058823529411765, 0.3215686274509804}
\definecolor{C4}{rgb}{0.5058823529411764, 0.4470588235294118, 0.7019607843137254}
\definecolor{C5}{rgb}{0.5764705882352941, 0.4705882352941176, 0.3764705882352941}
\definecolor{C6}{rgb}{0.8549019607843137, 0.5450980392156862, 0.7647058823529411}
\definecolor{C7}{rgb}{0.5490196078431373, 0.5490196078431373, 0.5490196078431373}
\definecolor{C8}{rgb}{0.8000000000000000, 0.7254901960784313, 0.4549019607843137}
\definecolor{C9}{rgb}{0.3921568627450980, 0.7098039215686275, 0.8039215686274510}

\usepackage{etoolbox}
\newcommand{\addQEDstyle}[2]{\AtBeginEnvironment{#1}{\pushQED{\qed}\renewcommand{\qedsymbol}{#2}}\AtEndEnvironment{#1}{\popQED}}

\newtheorem{thm}{Theorem}[section]
\newtheorem{lem}[thm]{Lemma}

\newtheorem{defn}{Defintion}[section]
\theoremstyle{remark}
\newtheorem{rem}{Remark}[section]
\addQEDstyle{rem}{$\triangleleft$}
\newtheorem{exa}{Example}[section]
\addQEDstyle{exa}{$\triangleleft$}

\usepackage{graphicx}
\graphicspath{{Figures/}}

\hypersetup{
    colorlinks,
    urlcolor=dbblue,
    citebordercolor=dbblue,
    linkbordercolor=dbblue,
    filecolor=dbblue,
    filebordercolor=dbblue,
    citecolor=dbblue,
    linkcolor=dbblue,
    colorlinks=true,
    pdftitle={RobustHedgingGANs},
    pdfsubject={},
    pdfauthor={Blanka Horvath, Yannick Limmer},
    pdfkeywords={hedging, robust finance, model uncertainty, knightian uncertainty, SigWasserstein-distance},
}

\begin{document}

\title{\textbf{Robust Hedging GANs} \\ {\normalsize Towards Automated Robustification of Hedging Strategies}}


\author[omi,math]{\href{https://orcid.org/0000-0002-8418-7284}{\includegraphics[scale=0.08]{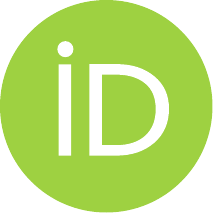}\hspace{1mm}Yannick Limmer}}
\author[omi,math]{\href{https://orcid.org/0000-0002-6369-7728}{\includegraphics[scale=0.08]{orcid.pdf}\hspace{.1mm} Blanka Horvath}}
\address[omi]{Oxford--Man Institute of Quantitative Finance}
\address[math]{Mathematical Institute, University of Oxford}

\date{This version: June 1, 2023.} 

\journal{TBA}

\begin{frontmatter}

\begin{abstract}
    The availability of deep hedging \cite{BuehlerGononTeichmannWood19} has opened new horizons for solving hedging problems under a large variety of realistic market conditions. At the same time, any model -- be it a traditional stochastic model or a market generator -- is at best an approximation of market reality, prone to model-misspecification and estimation errors.
    This raises the question, how to furnish a modelling setup with tools that can address the risk of discrepancy between anticipated distribution and market reality, in an automated way. 
    Automated robustification is currently attracting increased attention in numerous investment problems, but it is  a delicate task due to its imminent implications on risk management. Hence, it is beyond doubt that more activity can be anticipated on this topic to converge towards a consensus on best practices. 
    This paper presents a natural extension of the original deep hedging framework to address uncertainty in the data generating process via an adversarial approach inspired by GANs to automate robustification in our hedging objective. This is achieved through an interplay of three modular components:  
    \begin{enumerate*}[label=(\roman*)]
        \item a (deep) hedging engine,
        \item a data-generating process (that is model agnostic permitting a large variety of classical models as well as machine learning-based market generators), and
        \item a notion of distance on model space to measure deviations between our market prognosis and reality.
    \end{enumerate*}
    We do not restrict the ambiguity set to a region around a reference model, but instead penalize deviations from the anticipated distribution. Our suggested choice for each component is motivated by our aims for: Adaptability to generic path-dependent and exotic payoffs without major modifications of the setup, and applicability to highly realistic data structures and market environments, but other choices for each of the components are also possible.  We demonstrate this in numerical experiments to benchmark our framework against other existing results.
    Since all individual components are already used in practice, we believe that our framework is easily adaptable to existing functional settings. 
\end{abstract}

\end{frontmatter}


\section{Introduction}
The pursuit of generating more and more realistic market models has been on the agenda of quantitative financial modelling since the emergence of the Black-Scholes Model half a century ago.
Parallel to that, incorporating the possibility that models may misrepresent market reality in some way, or display other modelling errors in a dynamic and highly variable market environment, has been another major focus for researchers and financial analysts for decades \cite{Cont01}. More recently, especially since the emergence of deep hedging algorithms \cite{BuehlerGononTeichmannWood19}, Market Generation, or more precisely a neural-network-based synthetic generation of highly realistic market data \cite{BuehlerHorvathLyonsPerezWood20, HorvathIssaLemercierSalvi23, KondratyevSchwarz19, NiSzpruchWieseShujian20}, has become an active field of research as such flexible models are a key ingredient for a reliable performance on real data for deep hedging engines such as the one in \cite{BuehlerGononTeichmannWood19} and its later variants.  However, despite the new levels of flexibility that neural-network-based market models provide, recent modelling endeavours have led to a humbling realization that, 
it is unrealistic to expect perfect accuracy in capturing (or forecasting) the true underlying market dynamics at all times. As a result, automated robustification concepts are starting to attract increased attention in numerous investment problems including portfolio optimization \cite{BuehlerHorvathLimmerSchmidt23, Iyengar22} and hedging \cite{LutkebohmertSchmidtSester21,NiSzpruchSabatevidalesWieseXiaoShujian20,WuJaimungal23}, with each individual approach displaying their unique set of advantages and challenges, which we briefly reflect on in \Cref{sec:literaturereview} below. But robustification (especially in an automated setting) is  a delicate task due to its implications on risk management, which will no doubt continue to attract attention from researchers and practitioners alike, while our understanding of the related challenges matures into established best practices.

In the context relevant to us, modelling inaccuracies in capturing and predicting market dynamics can be broadly attributed to the following factors:
\begin{enumerate}[label=\roman*)]
    \item Markets are intrinsically heteroscedastic and non-stationary. Therefore, significant and lasting changes in market dynamics can manifest at any moment. 
    \item Especially when operating within small data environments (for instance characterized by a limited number of new input samples) fitting model parameters of a path generator is prone to estimation errors, leading to some degree of uncertainty about the underlying dynamics.
    \item The task of estimating future states is inherently challenging, even when focusing solely on predicting the most probable next step. Synthesizing the entire distribution forecast of future dynamics compounds the complexity of this task and be prone to known modelling errors.
\end{enumerate}
\medskip
To incorporate such possibilities of inaccuracies into the training of deep hedging engines efficiently, we have two aims in mind: We strive to update our beliefs and reference models in response to significant and lasting changes in market dynamics, but we aim to prevent minor estimation errors and moderate market fluctuations (fluctuations, which are limited in scale or time) from causing major disruptions to our algorithms. The latter is achieved by introducing robustification into the hedging engine, and the former can be detected algorithmically: See \cite{HorvathIssa23, HorvathIssaMuguruza21} for fast and efficient algorithmic methods to detect such regime changes and distinguish them from other market fluctuations and outliers.

As outlined above, this work is dedicated to showcasing a numerically efficient and model-free approach to address (various forms of) model uncertainty in a \emph{hedging context}, which is known to be an intricate challenge (see \cite{DolinskySoner14, NeufeldNutz13}), and we aim to do so in an (automated) algorithmic way by approximating corresponding optimal hedging strategies via a neural network. Our proposed method unites a powerful trio of recently established frameworks; 
    \begin{enumerate*}[label=(\roman*)]
        \item (deep) hedging engines,
        \item synthetic market data generation (possibly via generative models), and
        \item signature methods on path space; 
    \end{enumerate*}
to automate the generation of scenarios where there is uncertainty (ambiguity) about the true data generating process and the approximation of corresponding optimal strategies. In order to fuse these three components, we leverage the structures of a Generative Adversarial Network (GAN), that have emerged as one of the most prominent methodologies for generative modelling. However, our employment of GANs transcends the conventional objective of generating realistic synthetic data samples. Instead, we ingeniously harness the adversarial framework within GANs to bolster hedging strategies, reinforcing their resilience in the face of model uncertainty. By incorporating this framework, we strive to enhance the robustness of our hedging techniques, thereby ensuring their efficacy in navigating uncertain scenarios.

\subsection{Motivation for our modelling setup and the evolution of penalty-weighted worst-case approaches to model uncertainty}\label{sec:literaturereview}
Model uncertainty has been widely considered in the literature in several settings and using different terminologies. It is also referred to in terms of \textit{model ambiguity} or \textit{model-misspecification}, and under the umbrella of \emph{Knightian uncertainty} in Economics and of \textit{robust frameworks} in Finance. Model uncertainty generally be characterized in the following way: 
An agent is confronted with the task of evaluating the quality or desirability of a received payoff, which is represented as a random variable. To calculate this, the agent relies on the underlying probability distribution governing the payoffs. However, this distribution remains unknown or inaccessible to the agent. Consequently, the agent resorts to employing a family of possible models as a substitute for the unknown true distribution. As we outlined above, the agent's perception of the underlying distribution may deviate from the true distribution for various reasons.
 There is a rich body of literature modelling model uncertainty, ranging across several disciplines, and hence giving a full account of all scientific contributions to this area is beyond the scope of this section. One can, however, organize this body of literature by distinguishing two types of model uncertainty: \emph{Parameter uncertainty} and \emph{distributionally robust} approaches. Most available frameworks in the literature to date consider some form of parameter-uncertainty or parameter-misspecification \cite{CohenTegner19, LutkebohmertSchmidtSester21}, where a (stochastic) model is chosen to describe market dynamics, but its parameters are permitted to vary within a certain pre-specified range. Distributionally robust optimization (DRO) approaches \cite{BartlDrapeauOblojWiesel21, WuJaimungal23, Rahimian22} on the other hand---describe uncertainty about the model in a broader sense, i.e. permit distributional deviations from an anticipated reference distribution, beyond the scope of a parametric model. With the recent emergence of highly flexible market generators---some of which have been designed with the goal to reflect arbitrarily general (``data driven") market dynamics---this line between parametric and distributional model-misspecification becomes more blurred than in classical settings.
In this work we follow, strictly speaking, the first line of literature (parametric uncertainty); however, when we consider flexible generative models that can approximate arbitrary distributions, we can come close to distributional robustness. But in contrast to several DRO problems, our parametric setting, (combined with a penalization explained below) allows us to drop the frequently imposed restriction to $\epsilon$-balls around a reference measure.

\paragraph{Motivation for our modelling setup: Penalized worst-case approaches to model uncertainty}
One of the most prominent approaches to model uncertainty in the Economics literature---the worst case approach---considers a multitude of models simultaneously and evaluates the payoff under the worst possible choice of model. The conventional setting entails an agent maximizing their (expected) utility under the assumption that their model accurately describes the underlying system. When faced with uncertainty regarding the true model, the agent adopts a cautious stance by considering a worst-case scenario: The agent contemplates a situation where nature selects the worst-case model from a set of plausible models. Subsequently, the agent optimizes their utility based on this worst-case model, ensuring a strategy that can withstand adversarial conditions. Doing so is motivated by the work of \citeauthor{GilboaSchmeidler89}  \cite{GilboaSchmeidler89}, where they relax the axioms that were used in the seminal works of \citeauthor{NeumannMorgenstern07} and \citeauthor{Savage72} \cite{NeumannMorgenstern07,Savage72}\footnote{For a detailed discussion, refer to \cite{Schied06} and \cite[Section 5.2]{FoellmerSchied13}.}. 

Some come to the conclusion that there are a number of situations which may deem the use of worst-case approaches (in their original sense) too strict for real world applications, see e.g. \cite{FreySin99, BiaginiFrittelli04, Neufeld18, Bartl19}: The agent, after ensuring to withstand adversarial conditions even in the worst case, may end up with the trivial zero-strategy (i.e. do not invest in the risky asset) as a result, if  the corresponding robustification turns out to be ``too expensive'', or  ``too risky''.
Adding an appropriate penalty term may be a remedy in such cases: The authors of \cite{MaccheroniMarinacciRustichini06} suggest incorporating a penalty term into the objective function and thereby essentially placing a weighting on permissible competitor models which are regarded as less likely, or less relevant. It has been shown, that the  weighting of \citeauthor{MaccheroniMarinacciRustichini06} \cite{MaccheroniMarinacciRustichini06} combined with the duality of risk measures (cf. \cite{FoellmerSchied16}), is tantamount to maximizing the overall objective
\begin{align*}
    -\rho(u(X)) = \inf_{\bbP}\left(\myE{\bbP}{u(X)} + \alpha^{\mathrm{min}}_\rho(\bbP)\right),
\end{align*}
where $\inf_{\bbP}$ reflects the worst-case scenario that our risk-averse agent aims to withstand. Here, $u$ denotes a utility function and $\rho$ a convex risk measure, which by \citeauthor{FoellmerSchied16} implies the penalty $\alpha^\mathrm{min}_\rho$ for the competitor model $\bbP$.  \citeauthor{HansenSargent08} in \cite{HansenSargent08} were among the first to incorporate this type of uncertainty in the context of mathematical finance. One of its benefits is that in many cases the control problem can be approached by dynamic programming in the sense of \cite{Kalman63}. In a hedging context \citeauthor{HerrmannMuhleKarbeSeifried17} in \cite{HerrmannMuhleKarbeSeifried17} 
applied this form of model uncertainty to a generalized Black-Scholes-type setting where the volatility parameter can be time-dependent. By taking advantage of the observation that the problem can be addressed by dynamic programming, the authors arrive at an asymptotic characterization of the robustified problem, where the expansion is obtained in terms of small values of the parameter describing the agent's uncertainty aversion.

\paragraph{Our framework: A classical penalized worst-case approach to uncertainty, revisited with modern tools} Motivated by the initial success of \cite{HerrmannMuhleKarbeSeifried17} in a time-dependent Black-Scholes market, and motivated by the fact that today we have more powerful computational and algorithmic techniques at our disposal than what was available a half a decade ago, we 
also follow this line of research pioneered by \cite{MaccheroniMarinacciRustichini06,HansenSargent08}. However, we recast the worst-case optimization problem as an adversarial game and address it algorithmically via a GAN structure. We aim to approximate the \emph{robust} hedging strategies corresponding to this new optimization problems via neural networks rather than asymptotically, as was done in \cite{HerrmannMuhleKarbeSeifried17}. With these changes in place, we shall endeavour to allow for the possibility that the market model follows more general dynamics than in \cite{HerrmannMuhleKarbeSeifried17} (see details and examples in Section \ref{sec:Generator} types of path-generators considered). To accommodate the model-agnostic and potentially data-driven nature of the recast problem we will use as a penalty for deviations from our reference model, a model-agnostic signature MMD metric that has been successfully used in previous data-driven works to assess the similarity of two distributions on path-space to one-another.  Finally, instead of considering the classical expected utility operator, we apply a general utility operator, allowing for greater flexibility in modelling decision-making under uncertainty. The specific details of this generalization are expounded upon in the subsequent sections. Building upon this contextual foundation, and choice for our setting, we now proceed to formulate the optimization problem that our study aims to address.

As motivated in the paragraphs above, our optimization objective to derive optimal hedging strategies in a setting with model uncertainty formulated in a penalized worst-case approach to model uncertainty around a reference model reads as follows:
\begin{align}
   \min_{\phi \in \Phi} \ \ \left( \sup_{\bbP \in \calP} \myRM^\bbP \ \big((\phi \bdot \bfS^{\bbP})_T - \bfC_T\big) - \alpha(\bbP) \right),
    \label{id:HedgeProblem}
\end{align}
where $[0,T]$ denotes a time horizon for $T >0$, and $(\Omega, \calF)$ some measurable space. 
In the above,
\begin{enumerate}
    \item $\bfS^\bbP$ denotes a stochastic process with law $\bbP$, which takes values in $\R^d$, $d \in \N$. It denotes a possible model that describes the market dynamics, which may or may not coincide with our reference model (anticipated beliefs about market dynamics). Its law $\bbP$ belongs to some family of probability measures $\calP$ defined on some $\Omega$. As we will show, $\calP$ can be quite general\footnote{$\calP$ does not need to be  dominated nor do the measures contained in $\calP$  need be equivalent.}. We discuss details of $\bfS^\bbP$ and $\calP$ in \Cref{sec:Generator}.
    \item The strategy $\phi$ expresses the investment in 
    $\bfS^\bbP$ at times $0 \leq t_0 < \ldots < t_{N-1} < T$, $N \in \N$. Since this amount is held until the next trading time, the weighted increment $\phi_{t_{n-1}}\Delta \bfS_{t_n}$ is accrued in each time step $n \in \left\{ 1, \ldots, N \right\}$, hence we write
     $   (\phi \bdot \bfS)_T := \sum_{n = 1}^N \phi_{t_{n-1}}^\intercal \Delta \bfS_{t_{n}}$,
    where $\Delta \bfS_{t_{n}} = \bfS_{t_{n}} - \bfS_{t_{n-1}}$. 
    We give more details on the hedging strategy in \Cref{sec:DeepHedge}.
    \item The derivative contract to be hedged, is denoted by a terminal payoff $\bfC_T \in \calX$, where $\calX := \lbrace X: \Omega \to \R \rbrace$ denotes the set of real-valued random variables. We focus on vanilla payoffs here, but it is possible to generalize the problem to other option types. 
    \item The utility functional $\myRM^\bbP$ is defined as a mapping $\calX \to \R$ for every $\bbP \in \calP$, which is covered in \Cref{sec:Foundation}. One may think of it as a monetary risk measure. 
    \item The penalty function $\alpha: \calP \to \R$ describes the belief $\alpha(\bbP)$ that the agent has in the model $\bbP \in \calP$ and is usually represented as $\alpha(\bbP) := \gamma \, \alpha(\bbP, \bbQ)$, where $\gamma \in \R_+$ scales a distance notion $\alpha: \calP^2 \to \R_+$ from the anticipated (reference) model $\bbQ$. Requirements of $\alpha$ are discussed in \Cref{sec:Foundation}; and we introduce the $\sigw$-distance, as well as the $\sigmmd$, as an exemplary model agnostic distance in \Cref{sec:SigWMetric}.
\end{enumerate}

\begin{rem}[Interpretation of the robustified objective \eqref{id:HedgeProblem}]
    Let the utility operators $\myRM^\bbP, \bbP \in \calP$ correspond to monetary risk measures. Given some contingent claim $\bfX$, a common interpretation of the evaluation of $\myRM^\bbP(\bfX)$ is that it corresponds to the amount of cash to be held, that makes $\bfX$ acceptable from a risk perspective (under $\bbP$). With this, the penalty $\alpha(\bbP)$ corresponds to the reduction in this cash amount if $\bbP$ is an unlikely model. If there is  \textit{no uncertainty}, then only the cash requirement for the estimated model $\bbQ$ is calculated and optimized. 
\end{rem}

\subsection{The penalized worst-case approach as an adversarial neural network extension of the (deep) hedging problem} As outlined in the introductory sections, we address the optimization problem \eqref{id:HedgeProblem} via a generative adversarial network (GAN) structure. Where the roles for the discriminator and generator components are adjusted to the task at hand (see \Cref{sec:Generator}, \Cref{sec:DeepHedge} for detailed explanation). In this context, the generator assumes the adversarial persona of ``mother nature'', while the discriminator acts as a robust hedger, navigating through the suboptimal circumstances it confronts, metaphorically depicted as the ``lemons'' it is handed\footnote{This analogy draws inspiration from Dale Carnegie's renowned expression coined in his 1948 book, emphasizing the discriminator's resourcefulness in making the best of challenging situations.}.
The generator takes random noise as input and transforms it to data of the representing $\bfS$, which the discriminator then uses to calculate the utility of the corresponding hedges, i.e. the inner part of \eqref{id:HedgeProblem}. For this, the strategies $\phi$ will be approximated by a trainable deep neural network. As a start, observe that the discriminator corresponds in this setting to the classical deep hedge.  Then, the discriminator updates the network corresponding to $\phi$ to maximize the hedge utility, while the generator updates its trainable components to minimize the hedge utility, that we have penalized for the latter correspondingly by adding the penalty term.

\medskip
Note that analytic solutions for problems of type \eqref{id:HedgeProblem} have so far only been available for special cases, as the problem quickly becomes intractable for general settings. Some analytic and asymptotic results are available if sufficiently restrictive assumptions on the data generating process are imposed. We use these available results as benchmarks to verify our model in the numerical sections. For the more general settings, where benchmarks are not available, our algorithm permits to numerically obtain a solution for a hedging strategy (approximatively via a neural network) in a robust setting to overcome the limitations previously imposed by classical methods. The thus obtained approximative hedging strategies delivered by the network can be viewed as a candidate solution (to be verified in a separate step analytically if desired), for robust problems in far more general settings than what was attainable so far. 


\medskip
Our work can be seen as a natural extension of the (original) deep hedging framework of \cite{BuehlerGononTeichmannWood19} to tackle the robust problem. As a result, our approach retains the desirable features that characterize deep hedging, such as being model-agnostic and data-driven. More recently, various extensions and further developments of deep hedging methodologies have emerged, shifting the attention towards incorporating model uncertainty \cite{LutkebohmertSchmidtSester21, WuJaimungal23, GierjatowiczSabateVidalesSiskaSzpruchZurich20, Iyengar22, Ghahtarani22} into the setting. While all aforementioned contributions yield valuable insights for an efficient robustification of hedging strategies. We address the problem differently than the aforementioned ones, however in some cases with the appropriate choice of path generator and penalizer we can reconstruct some of these settings. Whenever possible, we compare our results with existing ones in the numerical sections.

The concept presented in \cite{LutkebohmertSchmidtSester21}, primarily focuses on handling parameter misspecification that assumes a uniform distribution of the model parameters between specified boundaries and subsequently optimizes over the combined measure. In contrast, our research diverges in that we do not presume any prior knowledge of the prevailing uncertainty. Instead, we allow the agent to select its own uncertainty aversion as a hyperparameter, facilitating seamless transitions between different model classes. It is important to note that this ansatz to uncertainty aligns with a problem formulation distinct from the one adopted in our paper. 
To assess performance of our approach in an out-of-sample test, we will employ a methodology similar to that employed by \cite{LutkebohmertSchmidtSester21}, as described in \Cref{sec:Numerics}.

Similarly to our study, \citeauthor{WuJaimungal23} focus in their recent work \cite{JaimungalPesentiWangTatsat22, WuJaimungal23}, on robustification outside the scope of parametric stochastic models. While their approach treats distributional robustness directly, we achieve this indirectly by specifying a general class of generators with universal approximation characteristics. 
The biggest difference between our approaches is that their notion of robustification alters the profit and loss directly via a neural network transformation within a Wasserstein ball, while we modify the underlying generator (and with that the paths that the strategy is tested on). In order to compare our respective results, the penalty term in our setting can restrict the profit and loss to a Wasserstein ball by a Dirac-type penalizer, which sets the penalty on the desired Wasserstein ball to zero and otherwise to infinity.  

Furthermore, our structure facilitates perturbations of the generator, irrespective of the resulting profit or loss. This attribute provides two notable advantages: Firstly, it eliminates the possibility that uncertainty is considered in the strategy. This becomes evident when considering model-free (or natural) hedges, which render robustification unnecessary. 
Secondly, it enables the evaluation of strategies under the same magnitude of model uncertainty regardless of their performance on the reference model. 


\citeauthor{BartlDrapeauOblojWiesel21} in \cite{BartlDrapeauOblojWiesel21} also incorporates Wasserstein balls, and their findings highlight the distinction between parameter uncertainty and distributional robustness. Notably, the authors achieve precise results. To facilitate their analysis, they consider asymptotically small Wasserstein balls and highlight aspects of directional changes as uncertainty is introduced.

This naturally leads us to the analytical result established by \citeauthor{HerrmannMuhleKarbeSeifried17} \cite{HerrmannMuhleKarbeSeifried17}, situated within the economic framework introduced \cite{MaccheroniMarinacciRustichini06} (which resembles our own economic framework the closest). Similarly as above, the authors had to adopt an asymptotic approach to render the problem mathematically tractable. Moreover, they encountered the need to confine their analysis to local volatility models, where the volatility component is assumed to be deterministic---an assumption that is not necessary in our work. As their setting constitutes a special case of ours, we are able to replicate their result, as demonstrated in \Cref{sec:Numerics}.

\medskip
In a more flexible "market generator" context, \cite{GierjatowiczSabateVidalesSiskaSzpruchZurich20} were the first to examine robustness in relation to neural SDEs, which constitute our baseline path-generator. For more on market generators, the reader may refer to \cite{KondratyevSchwarz19, BuehlerHorvathLyonsArribas20} as well as the work of \cite{CohenReisingerWang20, CohenReisingerWang21, CohenReisingerWang22}. In our work, we build upon some concepts derived in \cite{NiSzpruchWieseShujian20, NiSzpruchSabatevidalesWieseXiaoShujian20}, where market generators by the use of approximations of the $\sigw$-metric/$\sigmmd$ are first conceptualized. 

While robustness considerations are experiencing a surge of attention that can be traced back to use cases connected to deep hedging, it is worth noting that the notion of robustification goes back far further than that:  \citeauthor{Cont06} discusses already in 2006 \cite{Cont06} a framework similar to the one in \cite{MaccheroniMarinacciRustichini06}, which is defined via a premium for model uncertainty comparable to other risk measures and also allows an adversarial representation. Especially, the proposed framework is compatible with observations of market prices of a set of benchmark derivatives, which motivates the construction of our out-of-sample test cases (see \Cref{sec:Numerics}); where we are able to show exemplarily that the robust hedging GAN increases the performance of the deep hedge. 

\paragraph{Outline} The remainder of this article is organized as follows. We begin with a theoretical discussion of the problem, where we repeat the formulation of the problem in a mathematical context and examine properties of the loss function. Next, we will revisit the deep hedging framework, before we formulate the robust hedging GAN in detail. We conclude our analysis with numerical results, where we first replicate the result of \cite{HerrmannMuhleKarbeSeifried17} and then compare the out-of-sample performance of our approach against the classic deep hedge. We append a brief introduction to rough path theory, focussing on signatures and expected signatures, which are then used to derive the $\sigw$-distance; as well as some additional material relevant for the construction of the out-of-sample tests.

\section{Theoretical foundations}\label{sec:Foundation}
Let $[0,T]$ for some $T \in \R_+$ be the time horizon under consideration, and let $0 \leq t_0 < \ldots < t_{N-1} < T$, for some $N \in \N$ a collection of trading dates and $(\Omega, \calF, \frakF)$ a filtered measurable space, i.e. $\Omega$ is a state space and $\frakF := (\calF_t)_{t \in [0,T]}$ is a filtration with $\calF \supseteq \calF_T$ containing all market information. We assume that the market consists of $d \in \N$ hedging instruments with prices given by the $\R^d$-valued, $\frakF$-adapted stochastic process $\bfS$. In the following, we examine $\bfS$ under different measures $\bbP \in \calP$, where we write $\bfS^\bbP$ to emphasize the corresponding law. 
We assume without loss of generality that the risk-free interest rate is zero, i.e. our analysis is conducted on discounted values. For simplicity, we also assume the absence of transaction costs in this framework\footnote{However, all results in this work can be easily extended to transaction costs, as described in \cite{BuehlerGononTeichmannWood19}.}.

The set of possible trading strategies $\Phi$ is given by discrete $\frakF$-adapted processes, defined by $\phi := (\phi_{t_n})_{n \in \myset{1}{N}}$. It is possible to further restrict this set of admissible trading strategies, see for \cite{BuehlerGononTeichmannWood19} for more details.
\medskip

If our market is defined on the singleton $\calP = \lbrace \bbP \rbrace$, and is complete under $\bbP$, then there exists for any liability $\bfX^\bbP \in \calX$ a $\theta \in \Theta$ such that
\begin{align}
    -\bfX + p_0 + (\phi^{t_{n-1}} \bdot \bfS^{\bbP})_T \equiv 0. \label{id:Complete}
\end{align}
However, this is generally not the case, and therefore, the left side of \eqref{id:Complete} constitutes for every $\theta \in \Theta$ a random variable on $(\Omega, \calF, \bbP)$ which is referred to as \emph{profit and loss} (P\& L) of $\theta$ hedging $\bfX^\bbP$. With that, it is necessary to assign a numerical value to this random variable, to evaluate the success of a hedging strategy.
A popular choice uses \emph{convex risk measures}, see \cite{FoellmerSchied16}, by which one can quantify the success of $\phi \in \Phi$ hedging $\bfX^\bbP$ by a minimal amount of cash that, if added to the position, is sufficient to render the overall position acceptable\footnote{E.g. from the point of view of a supervising agency}. Let $\myRM^\bbP$ be a risk measure, then
\begin{align}
     \myRM^\bbP\big((\phi \bdot \bfS^{\bbP})_T - \bfX^\bbP \big) \label{id:RiskMeasure}
\end{align}
corresponds to this value.

Now, if $\calP$ contains more than one measure (i.e. there are more than one possible measures under which we may need to evaluate the risk measure), different $\bbP \in \calP$ may result in different numerical value of the capital needed: As we can see from the formula, \eqref{id:RiskMeasure} depends on the realized measure and the latter is a priori unknown to the agent.

Therefore, (under the worst-case approach) the agent computes the risk measure for each $\bbP \in \calP$ and performs the maximization for the worst of these scenarios.
To ensure, however, that the resulting evaluation is not too conservative, we lift the capital requirement partially by subtracting the penalty term $\alpha(\bbP)$. With that, we arrive at the optimization task \eqref{id:HedgeProblem}, defined by the following loss function
\begin{align}
    \calL: \Phi \times \calP \longrightarrow \R, \quad \phi \times \bbP \longmapsto \myRM^\bbP\big((\phi \bdot \bfS^{\bbP})_T - \bfC_T^\bbP \big) - \alpha(\bbP). \label{id:LossFunction}
\end{align}
\begin{thm}[Convexity]\label{thm:Convexity} 
    If $\myRM^\bbP$ is convex for all $\bbP \in \calP$, then $\sup_{\bbP \in \calP} \calL(\cdot, \bbP)$ is convex.
\end{thm}

For the sake of reading flow, the proofs of this section are delegated to the corresponding Appendix \ref{sec:Proofsofsection2}.

With regard to \Cref{thm:Convexity}, we are examining optimality under convex risk measures in the context of \cite{Xu06, KloeppelSchweizer07, IlhanJonssonSircar09}. To this end, we note that the optimal value of \eqref{id:HedgeProblem} is informative with respect to the efficiency to replicate a given contingent claim under model uncertainty. This entails a natural pricing framework, commonly referred to as \textit{indifference pricing}. For this, we first observe that the problem implicitly defines a convex risk measure.

\begin{thm}\label{thm:Convexitsecondtheorem}
Let $\myRM^\bbP$ be a convex risk measure for every $\bbP \in \calP$ and assume that the set of admissible strategies $\Phi$ is convex. Then, 
\begin{align*}
    \pi: \calX \longrightarrow \R, \quad  \bfX \longmapsto \inf_{\phi \in \Phi} \left\{\sup_{\bbP \in \calP} \left( \myRM^\bbP\big((\phi \bdot \bfS^{\bbP})_T + \bfX^\bbP \big) - \alpha(\bbP)\right) \right)
\end{align*}
is a convex risk measure. 
\end{thm}

Conclusively, $\pi(- \bfC_T)$ refers to the minimal amount of cash that, if injected in the hedging procedure, makes the position acceptable from a risk and uncertainty point of view, provided the position is hedged optimally. To obtain the indifference price, we have to normalize this cash-amount by the capital required for a zero-claim, i.e. 
\begin{align*}
    p(\bfC_T) := \pi(- \bfC_T) - \pi(0).
\end{align*}
We see that  $\pi(\bfC_T)$ actually coincides with the price of a replicating portfolio, if it exists.
\begin{lem}\label{lem:ThirdLemma}
    Suppose there exists a $\phi \in \Phi$ and $p_0 \in \R$ with 
    \begin{align*}
        \bfC^\bbP_T \equiv p_0 + (\phi \bdot \bfS^{\bbP})_T, \quad \forall \bbP \in \calP,
    \end{align*}
    then $p(\bfC_T) = p_0$.
\end{lem}
\begin{rem}[Arbitrage] 
The absence of arbitrage is not a requirement in our approach. Nonetheless, in accordance with the specifications in \cite{BuehlerGononTeichmannWood19}, we assert that certain markets can be considered irrelevant. In particular, this means that $\inf_{\phi \in \Phi}\rho((\phi \bdot \bfS^\bbP)_T) > -\infty$. It is important to note that market irrelevance does not solely stem from the presence of explicit arbitrage opportunities, as statistical arbitrage can also manifest itself in markets where arbitrage is not apparent. Furthermore, even if the market dynamics display clear-cut arbitrage opportunities and there are no constraints related to costs or liquidity, it does not guarantee our ability to exploit such opportunities. 
\end{rem}

We have readily established in \Cref{thm:Convexity} that the optimization task \eqref{id:HedgeProblem} is convex in the hedging strategy.
Also recall from \eqref{id:HedgeProblem}, that the optimization task also entails to find the supremum of $\calL$ with respect to $\bbP \in \calP$, which is substantially more difficult to characterize. If we, however, limit ourselves to OCE risk measures, we can prove concavity in the second component of the loss function $\calL$.

\begin{thm}[Concavity]\label{thm:Concavity}
    If $\myRM^\bbP$ is defined as an optimized certainty equivalent (OCE), i.e.
    \begin{align*}
        \myRM^\bbP(\bfX) = \inf_{w \in \R} w + \myE{\bbP}{u(-\bfX-w)}, \quad \forall \bbP \in \calP,
    \end{align*}
    where $u: \R \to \R$ is continuous, non-decreasing and convex, then $\calL(\phi, \cdot)$ is concave for every $\phi \in \Phi$ if $\alpha$ is convex. 
\end{thm}

The class of OCE risk measures is reasonably rich and includes the \textit{entropic risk measure} as well as the \textit{average value at risk} (also known as \textit{expected shortfall, conditional value at risk}), see \cite{FoellmerSchied16}.

\section{The data generating process (the path generator)}\label{sec:Generator}

As outlined in the introductory section, our setting is modular in all three of its components. To exemplify this modularity for the data generating process 
(and to demonstrate the versatility of our frameworks) we display some types of data generating processes\footnote{The permissible data generating processes are kept rather general, they permit a large class of transformations of a random noise to path-instances which correspond to possible simulations of the market. 
} 
that can can be applied in our general methodology. 
To remain faithful to the GAN nomenclature, we will refer to all the considered data generating processes as \emph{(path) generators}. We use the term path generators as a more inclusive term (entailing both classical and neural network models) than the notions of \emph{market generators} and \emph{market simulators}, which have been established\footnote{The terms \emph{market generators} and \emph{market simulators} have been coined to \emph{contrast neural network-based market models} from classical stochastic models} in the recent works \cite{BuehlerHorvathLyonsPerezWood20, KondratyevSchwarz19, NiSzpruchSabatevidalesWieseXiaoShujian20}. Despite staying faithful to the GAN nomenclature by using the term \emph{path generators}, let us note here that the role of the generative network in this context has shifted from its original primary goal of producing paths that resemble reality to a new goal of producing paths that challenge the hedger in a way determined by the modeller. 

To exemplify this, we state that path generators can be numerically formulated as mappings of the form
\begin{align}\label{eq:NSDE}
    \calG_\xi: 
    \begin{array}{ccc}
    \R^{N \times d'} & \longrightarrow & \R^{(N+1) \times d} \\
    (\Delta W_{t_n})_{n \in \lbrace 1, \ldots, N \rbrace} & \longmapsto & (S_{t_n})_{n \in \lbrace 0, \ldots, N \rbrace},
    \end{array}
\end{align}
transforming increments of a Brownian motion to a time-discretized path-instances of the asset price process.
We will denote by $\xi \in \Xi$ the parameters of this path generator, and by $\Xi$ the corresponding parameter set.\footnote{
 A path generator $\calG_\xi$ for a fixed parameter choice $\xi \in \Xi$ is a representation of some measure $\bbP$. Fix some Brownian increments $(\Delta W^J_{t_n})_{n \in \lbrace 1, \ldots, N \rbrace}$ for an index set $J \subseteq \N$. Then, a representation means that for all $n \in \myset{1}{N}$ the values $\calG_\xi((\Delta W^j_{t_n})_{n \in \lbrace 1, \ldots, N \rbrace})$ represent a sample path of $\bfS$ under $\bbP$ at times $(t_n)_{n \in \myset{1}{N}}$ for all $j \in J$.
} The modeller has the discretion in both---choosing $\mathcal{P}$, on the one hand, by specifying the type (and richness) of the model spaces included in $\mathcal{P}$, as well as any parameter restrictions imposed on it. The dimension $d' \in \N$ with $d' \geq d$ allows the formulation of hidden dimensions which are not tradable.

In the following, we list some choices of path generators:
\begin{enumerate}
    \item \textbf{Classical stochastic equity market models.} To observe cases that can be aligned with classical literature, one may choose any classical parametric equity market model or time series model as our path generating process specifying their parameters as \emph{``trainable"} (and possibly imposing suitable constraints on the space of admissible parameters). In these classical cases \emph{``training''} the parameters entails a standard calibration to market data or fitting to market observations.

    \item \textbf{NSDE-type generators.} For this general class of path generators, we assume that the measure $\bbP$ in $\calP$ is defined by arbitrary continuous maps $\mu^\bbP: [0,T] \times \R^{d'} \to \R^{d'}$, $\sigma^\bbP: [0,T] \times \R^{d'} \to \R^{d' \times d'}$ and thus dynamics of $\bfS$ given by 
    \begin{align}
                d\bfS_t = \mu^\bbP(t, \bfS_t)dt + \sigma^\bbP(t, \bfS_t)d\bfW^\bbP_t, \quad \bfS_0 = S_0 \in \R^{d'}, \quad \bbP \in \calP,
        \label{id:DynamicsRepeated}
    \end{align}
    and $\bfW^\bbP$ constituting a $\bbP$-Brownian motion with $d' \geq d$ components. The corresponding path generator can be obtained by sampling increments $(\Delta W_{t_n})_{n \in \lbrace 1, \ldots, N \rbrace}$ from $\bfW^\bbP$ and computing time-discretized samples of $\bfS$ via the implicit Euler-Mayurama scheme, i.e.
    \begin{align*}
        S_n = S_{n-1} + \mu^\bbP(t, S_{n-1})\Delta t_n + \sigma^\bbP(t, S_{n-1})\Delta W_{t_n}, \quad \forall n \in \left\{ 1, \ldots, N \right\},
    \end{align*}
    where $\Delta t_n := t_n - t_{n-1}$ and $\Delta W_{t_n} := W_{t_n} - W_{t_{n-1}}$. 
    To parametrize this type of path generator, the mappings $\mu^\bbP$ and $\sigma^\bbP$ can be described by deep neural networks parametrized by $\Xi^\mu$ and $\Xi^\sigma$. In this case, the overall parameter set becomes $\Xi := \Xi^\sigma \times \Xi^\mu \times \R^{d'}$\footnote{It may be advantageous to allow the initial condition of \eqref{id:DynamicsRepeated} to be specified by trainable parameters.}.
    To fully specify $\mathcal{P}$ in the problem \eqref{id:HedgeProblem}, parameter restrictions on $\mu^\bbP$ and $\sigma^\bbP$ can be applied\footnote{For instance, one can specify the absence of a drift (set $\mu^\bbP(\cdot, \cdot)^0 \equiv 0$) or the positivity of the volatility of the asset.}.
    \footnote{It is possible to transform the resulting samples after the initial transformation further, if for instance tradable volatility swaps should be available in the market. Eventually, it is possible to mask non-observable dimensions. }
    \item \textbf{LogSig-RNN.} In \cite{NiSzpruchSabatevidalesWieseXiaoShujian20}, the authors suggest to use, based on \cite{LiaoLyonsYangNi21}, to use a Log-Signature-based neural networks for generation purposes to obtain better performance with fewer parameters.
\end{enumerate}


To conclude this section, we denote the parametric version of \eqref{id:HedgeProblem}. For this, we assume that the payoff $\bfC_T$ is a function of $\bfS$, i.e. there is a function 
\begin{align*}
    c: ([0,T] \to \R) \longrightarrow \R, \quad S \longmapsto C_T,
\end{align*}
that determines the payoff structure of the contingent claim.\footnote{It is, in principle, possible to consider a more general class of derivatives. In order to achieve this, the payoff of the option has to be produced by the generator, alongside the asset prices.}
\begin{lem} \label{lem:Parametrization1}
    Let $\calP$ be represented by $\mySet{\calG_\xi}{\xi \in \Xi}$ and assume the operator $\myRM$ has a heuristic estimator $\hat{\myRM}$. Then, optimization problem \eqref{id:HedgeProblem} reads as 
    \begin{align*}
        \sup_{\xi \in \Xi} \hat{\myRM}\big((\phi \bdot S^I)_T - C_T^I\big) + \hat{\alpha}(\xi) \longrightarrow \min_{\phi \in \Phi} !,
    \end{align*}
    where $S^I := \lbrace \calG_{\xi}((\Delta W^i_{t_n})_{n \in \myset{1}{N}}) \rbrace_{i \in I}$ for corresponding parameters $\xi \in \Xi$ and sampled Brownian increments $(\Delta W^I_{t_n})_{n \in \myset{1}{N}}$ indexed by $I \subseteq \N$. Besides, $C_T^I := \hat{c}(S^I)$ with $\hat{c}$ being the heuristic version of $c$ and $\hat{\alpha}: \Xi \to \R$ is defined as $ \hat{\alpha}(\xi) = \alpha(\bbP)$ whenever $\calG_\xi$ represents $\bbP \in \calP$.
\end{lem}
\begin{proof}
     This follows directly from the definition of $\hat\myRM$, $\hat\alpha$, as well as the representation of $\calP$.
\end{proof}

\begin{rem}[Concavity of parametrized loss]\label{rem:Concavity}
    Note that the concavity established in \Cref{thm:Concavity} does not necessarily hold for the parametrized version using the path generator. 
 
    The task of identifying the global minimum in this problem can therefore pose a significant challenge, as it might involve navigating complex landscapes with numerous local optima. In such scenarios, the optimization process can easily become trapped in these local optima, failing to reach the desired global minimum. However, there exist various approaches to address this issue and still achieve satisfactory solutions\footnote{One commonly employed method is stochastic gradient descent (SGD), which plays a crucial role in overcoming the challenges associated with finding the global minimum.}. 
\end{rem}

\section{Deep Hedging}\label{sec:DeepHedge}
We briefly recall deep hedging \cite{BuehlerGononTeichmannWood19}\footnote{
\label{rem:DHE}
Note that we are using the basic deep hedging algorithm in this work, extensions include \cite{BuehlerMurrayPakkanenWood21, BuehlerMurrayWood22, MurrayWoodBuehlerWiesePakkanen22}.
}, since it is a substantial part of the proposed algorithm: One approximate the control $\phi$ -- formulated as a function of time and the then prevailing information to $\R^d$ -- via deep neural networks. For the sake of simplicity, we denote the approximated function by $\phi^\theta$, where $\theta \in \Theta$ describes the choice of weights in some parameter space $\Theta$. 
To fit the weights $\theta \in \Theta$ of this network, $\phi^\theta$ is tested on a batch of paths $S^{I}$, $I \subseteq \N$, sampled from $\bfS^\bbP$ that follows the (fixed) underlying model $\bbP$, by computing the heuristic utility
\begin{align}
    \hat{\myRM}\big( (\phi^\theta \bdot S^I)_T - C_T^I \big), \label{id:HedgeObjectiveHeursitic}
\end{align}
where $\hat{\myRM}$ is an estimator of the utility functional $\myRM$ and $C^I_T := \{C_T^{(i)}\}_{i \in I}$ corresponds to the payoff of the derivative on the sample path indexed by $I$. Then, the weights of the net approximating $\phi$ are updated by backpropagation to maximize \eqref{id:HedgeObjectiveHeursitic}.
We follow this approach as well to approximate hedging strategies, which will be expressed in the following Theorem.

\begin{thm} \label{lem:Parametrization2}
    Using the notation of \Cref{lem:Parametrization1}, then there exists for all $\epsilon > 0$ a parameter space $\Theta$ for neural networks such that 
    \begin{align*}
        \Big\vert \adjustlimits\inf_{\theta \in \Theta} \sup_{\xi \in \Xi} \hat{\myRM}\big((\phi^\theta \bdot S^I)_T - C_T^I\big) + \hat{\alpha}(\xi)  - \adjustlimits\inf_{\phi \in \Phi} \sup_{\bbP \in \calP} \calL(\phi, \bbP) \Big\vert < \epsilon.
    \end{align*}
\end{thm}
\begin{proof}
    The result follows directly from \Cref{lem:Parametrization1} with $\Phi = \mySet{\phi^\theta}{\theta \in \Theta}$, together with the approximation result of the deep hedge \cite[Proposition 4.3]{BuehlerGononTeichmannWood19} which is available due to \Cref{thm:Convexity}.
\end{proof}

\begin{rem}[Convexity of parametrized loss] \label{rem:Convexity}
    Computational considerations play a significant role in determining the (close-to) optimal parameters for neural networks. One common approach to finding a local minimum of a differentiable function is through the use of gradient descent algorithms. While the unparameterized loss function is convex in the strategy, this does not necessarily hold for the parametrized version. This is due to the fact that neural networks themselves are not convex, see \cite{Goodfellow16}.
    
    The crucial factor determining the success and feasibility of these algorithms is the ability to avoid getting trapped in local minima and instead converge towards the global minimum. To address this concern, the (minibatch) stochastic gradient algorithm has been devised, which mitigates the risk of being stuck in local minima. Nevertheless, it is important to note that, at present, there is no general theoretical guarantee ensuring convergence of neural networks to the global minimum within a reasonable timeframe. Nonetheless, there is a prevailing belief within the field that sufficiently large neural networks can attain a sufficiently low value of the cost function within a reasonable amount of time, as suggested by \cite{Goodfellow16}.
\end{rem}

\section{The robust hedging GAN}\label{sec:RobustHedgeGAN}
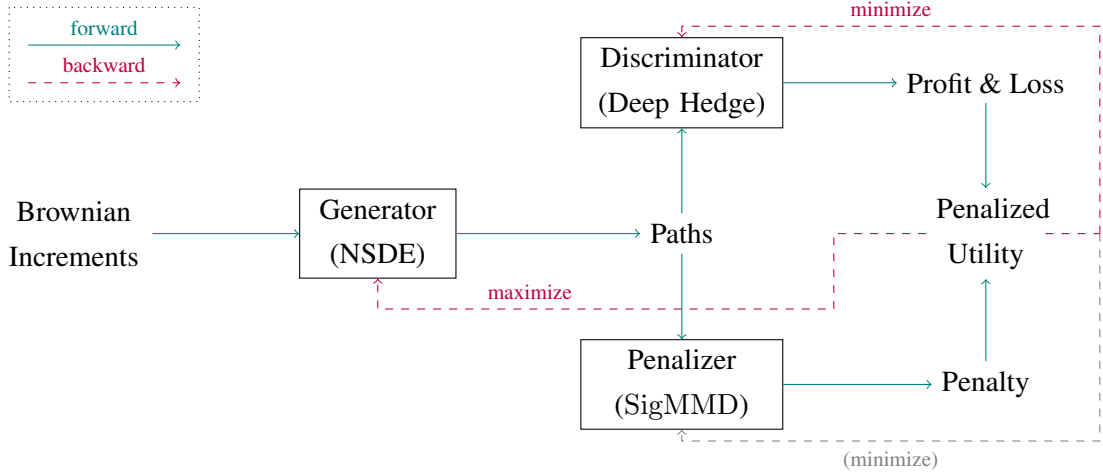
\begin{figure}
    \small
    \centering
        \begin{tikzpicture}
	        \node[text width=1.8cm, align=center] at (-4, 0) (gi) {Brownian Increments};
			\node[text width=1.8cm, align=center, draw] at (0, 0) (nsde) {Generator (NSDE)};
			\node at (4, 0) (model) {Paths};
			\node[text width=2.4cm, align=center, draw] at (4, 2) (dh) {Discriminator (Deep Hedge)};
			\node at (8, 2) (pnl) {Profit \& Loss};
			\node[text width=1.3cm, align=center] at (8, 0) (loss) {Penalized Utility};
			\node[text width=2.4cm, align=center, draw] at (4, -2) (pen) {Penalizer ($\sigmmd$)};
			\node at (8, -2) (penlty) {Penalty};
			
			\draw[->, teal] (gi) -- (nsde);
			\draw[->, teal] (nsde) -- (model);
			\draw[->, teal] (model) -- (dh);
			\draw[->, teal] (dh) -- (pnl);
			\draw[->, teal] (pnl) -- (loss);
			\draw[->, teal] (model) -- (pen);
			\draw[->, teal] (pen) -- (penlty);
			\draw[->, teal] (penlty) -- (loss);
			
			\draw[color=purple, dashed] (loss) -- (9.5,0);
			\draw[->, color=gray, dashed] (9.5,0) -- (9.5, -2.75) -- node[below, midway] {\scriptsize (minimize)} (4,-2.75) -- (pen);
			\draw[->, color=purple, dashed] (9.5,0) -- (9.5, 2.75) --node[above, midway] {\scriptsize minimize}  (4,2.75) -- (dh);
			\draw[->, color=purple, dashed] (loss) -- (6, 0) -- (6, -1) -- (4, -1) -- node[above, midway] {\scriptsize maximize} (0, -1) -- (nsde);
			
			\draw[->, teal] (-4.6,2.5) -- node[above, midway] {\scriptsize forward} (-2.6,2.5);
			\draw[->, purple, dashed] (-4.6,2) -- node[above, midway] {\scriptsize backward} (-2.6,2);
			\draw[dotted] (-4.85, 3) rectangle (-2.35, 1.75);
		\end{tikzpicture}
    \caption{Structure of the robust hedging GAN with backpropagation of the components.}
    \label{fig:Structure}
\end{figure}

\Cref{lem:Parametrization2} formulates the hedging problem under uncertainty \eqref{id:HedgeProblem} as a parametric problem. We now suggest constructing the \emph{robust hedging GAN} as a solution to this optimization task. The structure consists of a generator-discriminator pair that has been presented individually in \Cref{sec:Generator}, \Cref{sec:DeepHedge} together with a penalizer (see Appendix \ref{sec:Distances}). We now briefly discuss their interplay, which is visualized in Figure \ref{fig:Structure}.
\begin{enumerate}
    \item \textbf{Generator.} The generator $\calG_{\xi}$, $\xi \in \Xi$ receives sample paths of a Brownian motion and transforms them into increments of a path-realization of the asset price process $\bfS^\bbP$ under the model $\bbP$ that is represented by $\bbP$. 
    \item \textbf{Discriminator.} The discriminator corresponds to the deep hedge instance. For the generated batch of paths, it computes via the strategy $\phi^\theta$ that is defined by its neural network weights $\theta \in \Theta$ the respective profits and losses and interprets them with the estimator of the utility functional $\hat{\myRM}$, as in \eqref{id:HedgeObjectiveHeursitic}. 
    \item \textbf{Penalizer.} The penalizer now penalizes the model $\bbP$ which is expressed in form of the generator $\calG_\xi$ with its parameter choice $\xi$. The user can specify any desired custom penalty term, however, we suggest using the $\sigmmd$ (in particular an approximation thereof), a metric that directly operates on path space. For details, we refer to Appendix \ref{sec:Distances}. With that, the penalty term for a path generator in question is determined by generating a batch of paths from it and measuring the distance to a batch of reference paths\footnote{Sometimes the reference data is not in the form of paths (as it is for instance the case with volatility smiles). In these cases, we calibrate a path generator to the reference data and use a batch of paths generated from it as reference paths. As we will point out below, it is helpful to use the reference path generator as an initialization to the adversarial training phase---therefore no additional computational cost arises.} and therefore is \emph{agonistic} to the choice of path-generator. Note that in our case, the penalizer does not inherit trainable variables, but it is possible to allow this and therefore incorporate penalty terms that are defined via a supremum. 
\end{enumerate}

When it comes to fitting the optimal parameters $\theta \in \Theta, \xi \in \Xi$, it is beneficial to pre-train the individual components on a standalone basis. This is done in the following procedure.
\begin{enumerate}
    \item Define and calibrate the path generator to establish a data generating process, representing the reference model $\mathbb{P}$. The calibration process aims to align the state $\xi \in \Xi$ of the path generator to past observations of historical data or some option-induces market state.
    
    As we mentioned in \Cref{sec:Generator}, the data generating process can constitute either a parametric equity market model such as Black-Scholes, Heston, Rough Bergomi, etc.; or a modern machine-learning based market generator. The initial choice not only establishes our belief regarding the current distribution $\mathbb{P}$ of the market, but also defines the possible directions $\calP$ in which the market can evolve. Specifically, it determines the number of parameters and their specific values that can vary under model uncertainty.
    \item Fit the deep hedge component\footnote{At this point, this is tantamount to a standalone deep hedge (without model uncertainty) as first presented in \cite{BuehlerGononTeichmannWood19}.} to the trained generating process from the previous step, by repeatedly generating paths from the generator and optimizing the trainable weights of the deep hedge such that \eqref{id:HedgeObjectiveHeursitic} is maximal.
\end{enumerate}
Note that the combination of the above steps is tantamount to an application of the deep hedge without awareness for uncertainty. However, it might be the case, that our initial calibration step derived a generator that represents a faulty model $\bbQ$. Possible scenarios where this occurs include the following cases:
\begin{enumerate}
    \item The market is non-stationary and the true underlying distribution $\mathbb{P}$ has moved away from the reference model $\mathbb{Q}$,
    \item the calibration is performed under simplifications, such as static implied volatility values,
    \item the generator is over-parametrized and therefore overfitted to the available data,
    \item the actual dynamics can not be represented by an overly simplified generator class, or
    \item the calibration of the generator is imprecise due to a lack of data.
\end{enumerate}

For this reason, we add an adversarial training procedure, which defines the robust hedging GAN. For this, the above described forward pass is executed:
\begin{enumerate*}[label=(\roman*)]
    \item the generator, defined by $\xi$, creates path instances, which are 
    \item then used to apply the hedging strategy, defined by $\theta$ and 
    \item are interpreted as a utility value by applying $\hat\myRM$. In a separate step, the generated paths are used to compute the penalty term.
\end{enumerate*}
    
To this end, we update the parameters $\xi, \phi$ iteratively. If we update $\xi$ in this step, we are trying to maximize the utility value (which is interpreted as loss). From an economical point of view, we let the generator move away from our highest belief and explore nearby areas where the hedge performs the worst. To limit the exploration of the generator in this step, the penalty term decreases the loss if the generator moves away from the agent's highest belief.

Afterwards, we update the parameter $\theta$ which has the opposite goal, it tries to minimize the loss. Loosely speaking, the deep hedge adapts to the latest change of the generator. Clearly, the strategy is still required to perform adequately on the previous generator states, otherwise the generator will simply step back -- encouraged by the penalty function. With this, the strategy is forced to become more robust.  

An implemention of the \textit{robust hedging GAN} may be found on GitHub\footnote{\url{https://github.com/YannickLimmer/adversarial-robust-deep-hedging}}; which includes the numerical results of this work.

\begin{rem}[Convergence]
    \Cref{thm:Convexity} and \Cref{thm:Concavity} indicate that the GAN iterates on a parametrized version of a convex-concave function. In general, however, we can not state that the parametrized version (see \Cref{lem:Parametrization2}) is convex-concave, see \Cref{rem:Concavity} and \Cref{rem:Convexity}. With this, we cannot exclude the possibility of the GAN to converge to local optima (cf. \cite{Kodali17}).

    We implement two effective strategies to mitigate this limitation. Firstly and as argued before, we employ a stochastic descent algorithm, significantly reducing the chances of being trapped in local extrema. This approach enhances the overall flexibility and adaptability of the system. Secondly, we incorporate pre-training of the deep hedge and path generator prior to the adversarial training phase. Although the resulting local extremum may not achieve the exact desired level of robustness, it still contributes to fortifying the deep hedge in a meaningful manner.
\end{rem}

\section{Numerical results} \label{sec:Numerics}
In this section, we will demonstrate the usage of the \emph{robust hedging GAN}. We will conduct three experiments, \begin{enumerate*}[label=(\roman*)]
    \item we reproduce the results of \cite{HerrmannMuhleKarbeSeifried17} in a exemplary Black-Scholes setting, 
    \item we investigate its out-of-sample performance in an artificial model uncertainty setup within the Black-Scholes model, and eventually 
    \item move on to an out-of-sample performance test based on the Heston model and motivated by real market data.
\end{enumerate*}  

\subsection{Comparison to \texorpdfstring{\citeauthor{HerrmannMuhleKarbeSeifried17} \cite{HerrmannMuhleKarbeSeifried17}}{}} 

We begin with a simple example, the Black-Scholes model \cite{BlackScholes73}. In the arbitrage-free case -- with drift parameter zero -- this model has a single parameter, the volatility.
The generator in this case is of the form 
\begin{align}
    d\bfS^{\bbP}_t = \sigma^{\bbP} \bfS^{\bbP}_t d\bfW^{\bbP}_t, \quad t \in [0,T] \quad \text{and} \quad \bfS^{\bbP}_0 \equiv S_0,
    \label{id:BlackScholesReference}
\end{align}
where we use the notation of \eqref{id:DynamicsRepeated} and specify $\sigma^{\bbP} \in \R$ as well as $\bfW$ as a Brownian motion under every choice of $\calP$. In particular, we assume that $\calP$ is the set of measures $\bbP$ such that dynamics \eqref{id:BlackScholesReference} hold for some $\sigma^\bbP \in \R^+$. Since a generator representing $\bbP$ is fully defined by $\sigma^\bbP$, we can define $\Xi := \R_+$. 

For our example, the estimated model $\bbQ$ is defined via $\sigma^{\bbQ} := 0.2$, and we allow trading on $18$ time-steps with step-size $5/255$, what corresponds approximately to a week in a business day calendar. We set $S_0 = 1$ and, therefore, fix for the hedging problem an at-the-money European call option with strike price equal to $1$, what results in a payoff $\bfC_{T} = (\bfS_{T} - 1)_+$. 

Since \cite{HerrmannMuhleKarbeSeifried17} use expected-utility operators, we set
\begin{align*}
    \myRM^\bbP(\bfX) := \myE{\bbP}{(1 - \exp(- \lambda \bfX)) / \lambda}.
\end{align*}
With this, we have everything specified for the standalone deep hedge application, which is carried out with two hidden layers of $128$ nodes each, fitted with ``ReLu'' activations and an identity output activation.

For the robust hedging GAN, we fix the penalty as\footnote{In the notation of \cite{HerrmannMuhleKarbeSeifried17}, this means we fix \begin{align*}
    f(t, s, y; \sigma) := (\sigma - \bar{\sigma})^2 / T.
\end{align*}} 
\begin{align*}
    \alpha_\phi(\bbP) := \frac{1}{\gamma N}\myRM^\bbP(\PnL^\phi)\sum_{n=1}^N\left( \sigma^\bbQ - \sqrt{\frac{\myV{\bbP}{\log(\Delta\bfS_{t_n})}}{\Delta t_n}}\right)^2.
\end{align*}

\begin{rem}[Homogeneous penalty] 
    The authors \citeauthor{HerrmannMuhleKarbeSeifried17} provide several reasons for including the hedge objective in their penalty. First, it ensures that preferences are invariant under affine transformations of the utility function. Second, incorporating this utility value allows for a dynamic formulation of the hedging problem, taking into account the initial time, stock price, and profit and loss, which eases the effort to derive an analytical solution. Finally, the authors' results suggest that if this term is not present, the optimal strategy is dependent on the prevalent profit and loss to formulate the hedging strategy at each time. 

    It is clear that our framework (as formulated in \Cref{sec:RobustHedgeGAN}) can include a relation to the hedge objective in the penalty term, and we will adopt this setting for this example. The reader may refer to the implementation of this example for more details. Nonetheless, we want to highlight that an inclusion of the hedge objective is not a necessity in our framework, and we argue that it entails benefits if the uncertainty analysis refers to the asset alone and is independent of the considered strategy. 
\end{rem}

\begin{figure}
    \centering
    \includegraphics[width=.8\textwidth]{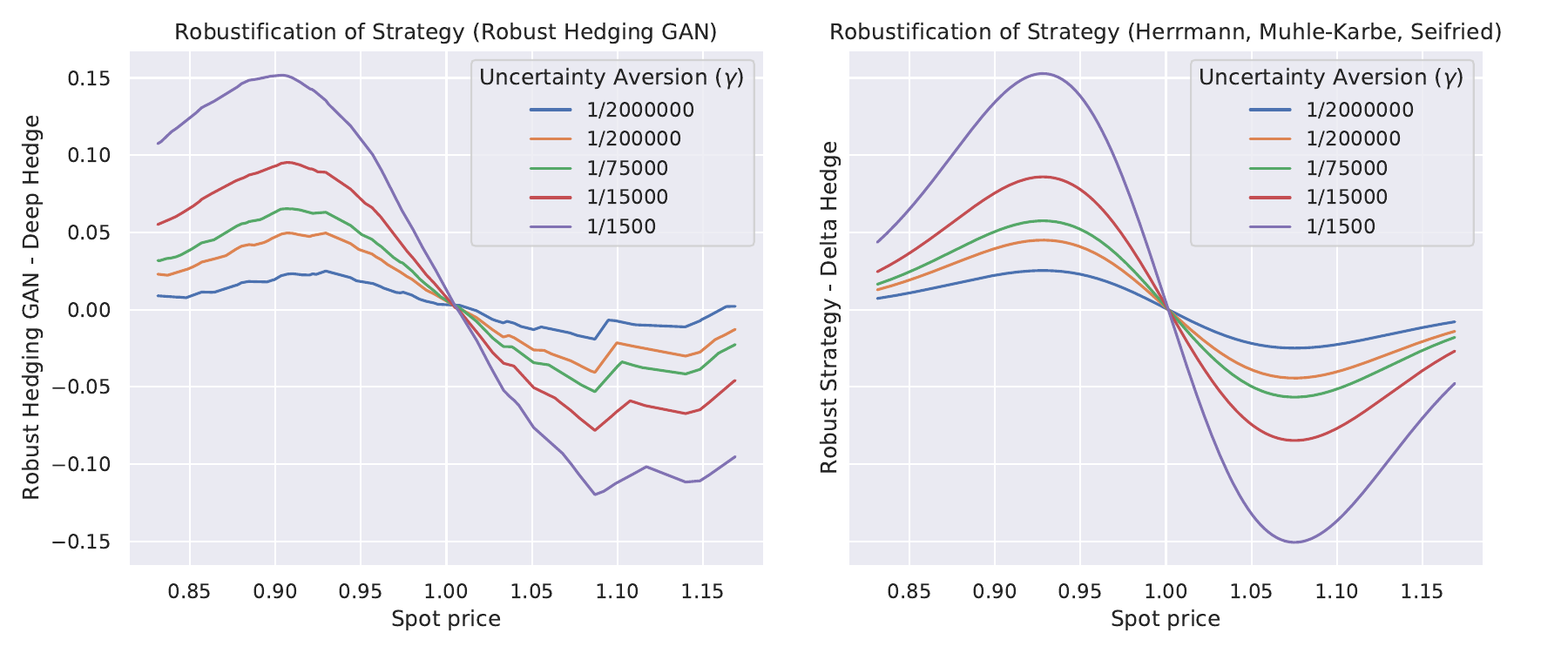}
    \caption{Comparing the robust hedging GAN strategy to the results of \cite{HerrmannMuhleKarbeSeifried17}.}
    \label{fig:ComparisonHMS}
\end{figure}

The authors \citeauthor{HerrmannMuhleKarbeSeifried17} prove that for small uncertainty-aversions $1/\gamma$ the optimal strategy is given by
\begin{align*}
    \delta^\gamma: 
    \begin{array}{ccl}
        [0, T] \times \R_+ \times \R & \longrightarrow & \R \\
        (t, s, y) & \longmapsto & \delta^{\mathrm{BS}}(t, s) + \gamma \delta^{\mathrm{U}}(t, s, y)
    \end{array}
\end{align*}
where $\delta^{\mathrm{BS}}(\cdot, \cdot)$ is the Black-Scholes delta and the uncertainty correction $\delta^{\mathrm{BS}}(\cdot, \cdot)$ is defined as 
\begin{align*}
    \delta^{\mathrm{U}}: 
    \begin{array}{ccl}
        [0, T] \times \R_+ \times \R & \longrightarrow & \R \\
        (t, s, y) & \longmapsto & \frac{\partial}{\partial s}w(t, s, y), 
    \end{array}
\end{align*}
where the source function $w: [0, T] \times \R_+\times \R \rightarrow \R$ which is given by the following PDE 
\begin{align*}
\frac{\partial}{\partial t} w(t, s) + \frac{1}{2}\sigma^2 s^2 \frac{\partial}{\partial s^2} w(t, s) + \frac{T\big(\sigma^\bbQ s^2 \gamma^{\mathrm{BS}}(t, s)\big)^2}{4} &= 0, \quad \forall t, s \in [0, T] \times \R_+, \\
w(T, s) &= 0, \quad \forall s \in \R_+.
\end{align*}
Hereby, the term $\gamma^{\mathrm{BS}}(\cdot, \cdot)$ refers to the Black-Scholes gamma. The term $\delta^{\mathrm{U}}$ is visualized for $t = T/2$ in the right-hand side of \Cref{fig:ComparisonHMS}.

We opposed this graph with the corresponding difference of the robust hedging GAN strategy and the deep hedge strategy at this point in time. First, we observe that both figures mostly coincide. With that, we obtain an insight on how this economic framework robustifies the strategy: 

The adjustment of the strategy is independent of the current profit and loss and the shape of the utility function. When the option is out of the money, an increase in the stock price increases the vulnerability of the replication strategy to volatility misspecification. However, the uncertainty-aware strategy, which holds more shares than a pure delta hedge, compensates for this increased vulnerability by generating additional profit. Conversely, when the stock price decreases, the strategy adjustment results in a loss compared to the pure delta hedge, but this loss is balanced by the fact that the stock price moves to a region with lower vulnerability to volatility misspecification. We can observe analogue dynamics with opposite direction in the case of in the money options. In summary, the almost optimal strategy serves as a hedge against movements of the stock price into regions with high vulnerability to volatility misspecification. A conclusion by \citeauthor{HerrmannMuhleKarbeSeifried17} is that the adjustment is primarily determined by the option's cash gamma, the almost optimal strategy can be understood as a hedge against movements of the stock price into regions with high cash gamma.

\begin{rem}
There are multiple reasons for minor differences in the two graphs in \Cref{fig:ComparisonHMS}:
\begin{enumerate}
    \item The roughness observed in the left graph can be attributed to two main factors. Firstly, the neural networks employed in the model architecture are relatively simple, comprising only two layers with 128 nodes each. This limited network complexity can contribute to the coarseness of the resulting graph. Secondly, the graph represents the disparity between two states of the same deep hedge, which amplifies the visibility of the edges and leads to a more jagged appearance.
    \item In our model setup, the trading frequency is set to occur every 5 business days. This infrequent trading interval creates a non-complete market scenario, causing the deep hedge to deviate from the pure Black-Scholes delta hedge strategy. 
    \item On the right-hand side, the analytical solution derived exhibits only asymptotic optimality. This outcome is a consequence of employing a quadratic Taylor expansion, which inherently introduces certain limitations to the accuracy and precision of the solution. Therefore, while the derived analytical solution provides some level of optimality, its effectiveness is somewhat constrained, and further improvements may be achievable through alternative approaches.
\end{enumerate}
\end{rem}
\begin{rem}[Comparison to further existing results]
    As mentioned in the introduction, we refer to the article \cite{WuJaimungal23}, where the authors have also obtained findings consistent with our own research. Recall that their approach alters the profit and loss directly via a neural network transformation within a Wasserstein ball. As they explore the hedging of knock-in call options, while our study focuses on vanilla options, recall that the instruments are equivalent once knocked in. It is worth noting that although we utilize a penalty different from theirs, we arrive at a parallel outcome; which is qualitatively described above in the comparison to \cite{HerrmannMuhleKarbeSeifried17}.\footnote{To see the similarity of the robustification, compare the left-hand side of \Cref{fig:ComparisonHMS} with Figure 3 (first column, middle) in the corresponding article.}

    In this specific example, there is one crucial aspect that remains undetectable in our analysis. It pertains to the concept of distributional robustness, which is successfully achieved by the model-free Wasserstein ball, as realized in \cite{WuJaimungal23}. However, it should be recalled that our analysis here is based on the replication of the findings from \cite{HerrmannMuhleKarbeSeifried17}, which primarily focuses on parameter uncertainty. The detailed explanation of the dissimilarity between the two approaches can be found in the comprehensive work by \cite{BartlDrapeauOblojWiesel21}.

    We hypothesize that by selecting a flexible enough generator class (for examples, refer to \Cref{sec:Generator}), we will be able to mimic distributionally robust results. This will be discussed in more detail in \Cref{sec:NSDEApplication}.

\end{rem}

\subsection{Out-of-sample test in the Black-Scholes model} 

As a next step, we want to examine the out-of-sample performance of the robust hedging GAN and compare it to the deep hedge. For this, we continue in the above Black-Scholes setting, however define the penalizer as a comparison of the volatility estimates of the generated paths $\sigma^\bbP$ to the volatility estimate of the batch $\sigma^\bbQ$, weighted by $\gamma > 0$. This means, 
\begin{align*}
    \alpha(\bbP) :=  \frac{1}{\gamma N}\sum_{n=1}^N\left( \sigma^\bbQ - \sqrt{\myV{\bbP}{\log(\Delta\bfS_{t_n})}}\right)^2.
\end{align*}

\begin{exa}[Black-Scholes test case]\label{exa:BST}
    We assume that the derivative, trading frequency, initial asset price, set of eligible parameters $\Xi$, and estimated parameter $\sigma^\bbQ$ are specified as above.

    The hedging objective in this example is chosen as the \textit{entropic risk measure}, i.e.
    \begin{align*}
        \myRM^\bbP(\bfX) := \frac{1}{\lambda}\log\myE{\bbP}{\exp(- \lambda\bfX)}
    \end{align*}
    for any random variable $\bfX$, where we fix the risk aversion as $\lambda = 130$.\footnote{This choice can roughly be motivated by considering mean-volatility, i.e. $\myRM^\bbP(\bfX) := \myE{\bbP}{\bfX} - \frac{\lambda'}{2}\sqrt{\myV{\bbP}{\bfX}}$, where we manipulate the expectation to the lower bound of the confidence interval given by $\lambda'/2$ standard deviations. A reasonable choice for the mean-variance risk-aversion is thus $\lambda = \lambda'/ V$ with $V^2 \approx \myV{\bbP}{\bfX}$. We know that the entropic risk measure coincides with the mean-variance operator for normally distributed random variables.}
\end{exa}

\begin{figure} 
    \centering
    \includegraphics[width=.6\textwidth]{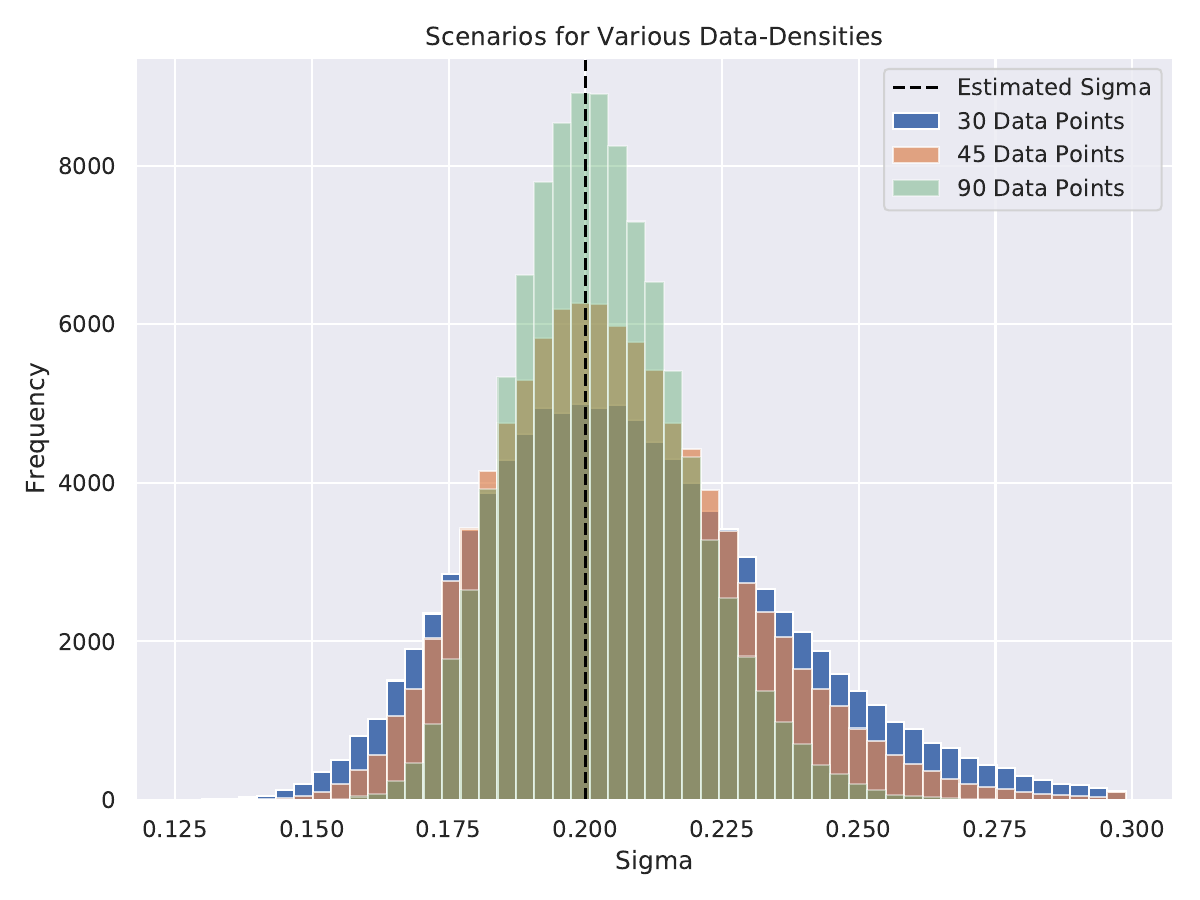}
    \caption{Sigma parameters inverse calibrated to the estimated model with various data densities.}
    \label{fig:UncertaintySigmas}
\end{figure}

\begin{exa}[Black-Scholes out-of-sample test data]\label{exa:CalibrationBS}
    The out-of-sample test data in this case should shadow the following situation. An agent assumed that the Black-Scholes model (with zero drift) is a suitable model to portray market dynamics, so she chooses the volatility parameter $\sigma^{\bbQ}$ corresponding to the realized volatility on historical asset price data. In order to have a plausible stationarity assumption for the estimation of $\sigma^{\bbQ}$, she only considers the last $45$-business days of data. 

    Now, let us assume that the agent chose $\sigma^{\bbQ} := 0.2$ and we construct our test set by asking---what is the volatility parameter in the true underlying (Black-Scholes) model, and what is the corresponding probability? Since we know the size of the data sample, we are in fact able to reverse-engineer this distribution. We showcase this distribution for 30/45/90 data points in \Cref{fig:UncertaintySigmas}.

    The robust hedging GAN from \Cref{exa:BST} can now be tested on various ``true underlying volatility values'' and eventually statistically evaluated. This type of backward test is explained in great detail in \Cref{sec:Test}.
    
    To test the strategies, we compute the out-of-sample loss (see \eqref{id:OOSP2} for details) with $M = 10\,000$ where we approximate the utility by sampling $10\,000$ instances of the true underlying model $\bfS^{\bbP^m}$ for each $m \in \lbrace 1, ..., M \rbrace$ respectively.
\end{exa}

\afterpage{
\begin{table}[p]
    \centering
    \footnotesize
    {\setlength{\tabcolsep}{16pt}
    \begin{tabular}{cccc}
        \toprule
        \textbf{Strategy} &  $1/\gamma$ & \textbf{Mean} & \textbf{Standard Deviation} \\ \midrule
        Deep Hedge       &                                       & \texttt{0.055361}  &                 \texttt{0.007137}  \\ \midrule
        \multirow{7}{*}{robust hedging GAN} &  10 & \textcolor{C1}{\texttt{0.056382}} &  \textcolor{C2}{\texttt{0.005982}} \\
         &  25                                    & \textcolor{C1}{\texttt{0.055471}} &  \textcolor{C0}{\texttt{0.006417}} \\
         &  50                                    & \textcolor{C0}{\texttt{0.055315}} &  \textcolor{C0}{\texttt{0.006674}} \\
         &  75                                    & \textcolor{C0}{\texttt{0.055286}} &  \textcolor{C0}{\texttt{0.006815}} \\
         & 100                                    & \textcolor{C2}{\texttt{0.055271}} &  \textcolor{C0}{\texttt{0.006899}} \\
         & 125                                    & \textcolor{C0}{\texttt{0.055286}} &  \textcolor{C0}{\texttt{0.006927}} \\
         & 150                                    & \textcolor{C0}{\texttt{0.055299}} &  \textcolor{C0}{\texttt{0.006986}} \\ \midrule
        Hedge on Test Set   &                     &                \texttt{0.055278}  &                 \texttt{0.006860} \\ \bottomrule
    \end{tabular}
    }
    \caption{Mean and standard deviation of out-of-sample performances (loss) of the test performed in Example \ref{exa:BST} for various penalty scales.}
    \label{tab:BlackScholesExample}
\end{table}
\begin{figure}[p]
    \centering
    \includegraphics[width=.49\textwidth]{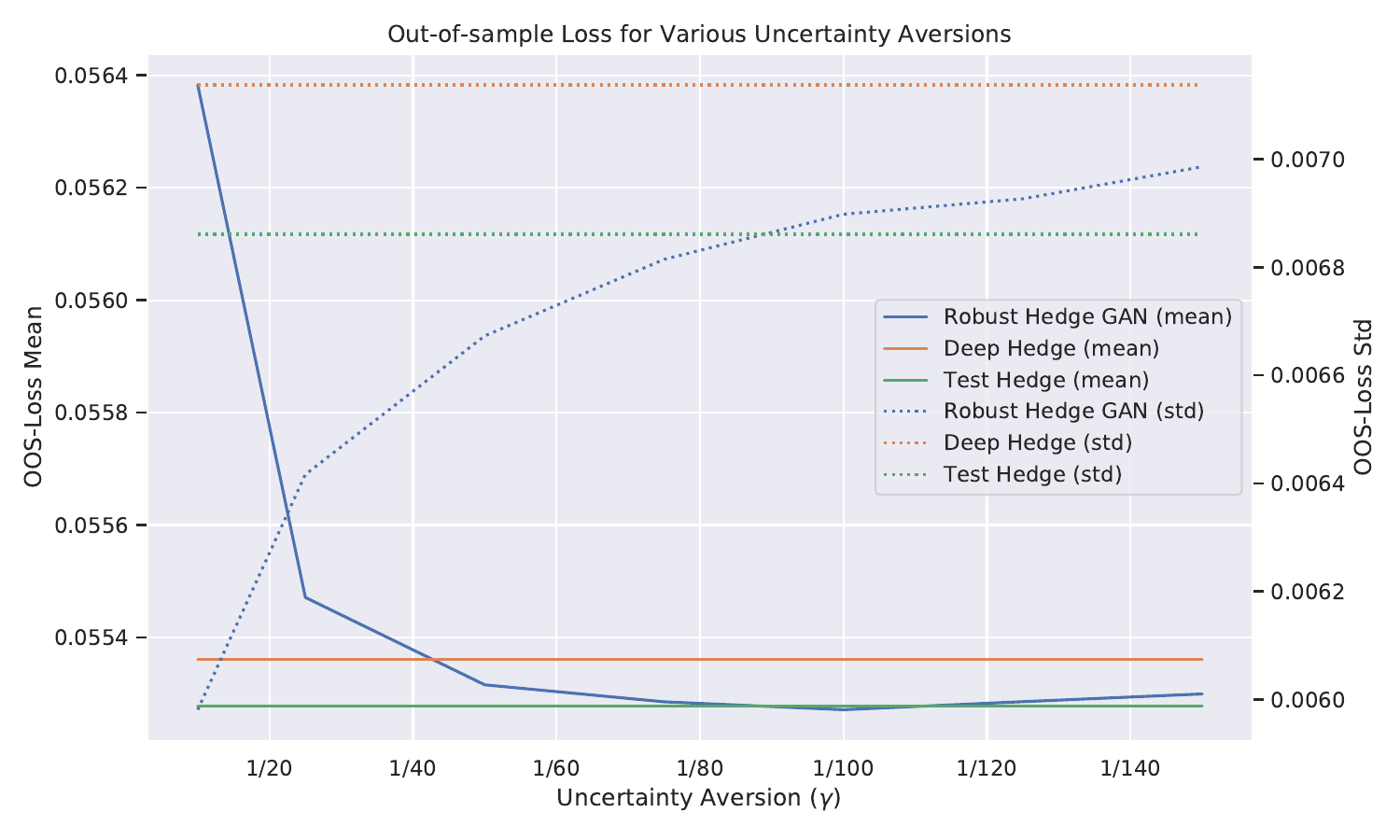}
    \caption{Mean and standard deviation of out-of-sample performances (loss) of the test performed in Example \ref{exa:BST} for various penalty scales.}
    \label{fig:BlackScholesExamplePens}
\end{figure}
\begin{figure}[p]
    \centering
    \includegraphics[width=.99\textwidth]{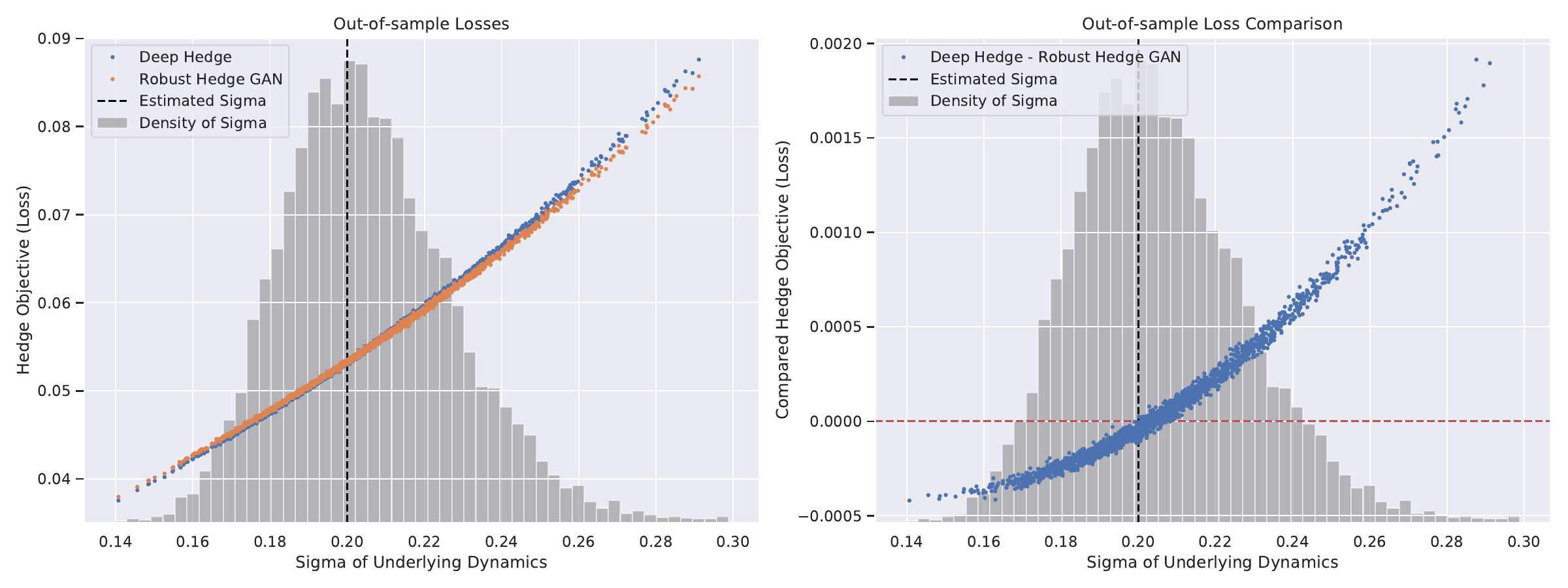}
    \caption{Out-of-sample performances (loss) for individual scenarios of the test performed in Example \ref{exa:BST} with $\gamma := 1/75$.}
    \label{fig:BlackScholesExampleDetail}
\end{figure}
\medskip
}

The estimated mean and variance of the resulting out-of-sample performances is displayed in \Cref{tab:BlackScholesExample} and visualized in \Cref{fig:BlackScholesExamplePens} for various choices of $\gamma$. It is evident, that the out-of-sample loss can in average be decreased by applying a robust hedging GAN instead of the standalone deep hedge. This is in particular the case for lower values of $\gamma$, with the lowest value resulting from the choice $\gamma = 1/100$.

In general, we see that the value converges to the deep hedge for decreasing uncertainty aversion, which aligns with our understanding that deep hedge and robust hedging GAN coincide for no uncertainty aversion. When increasing the uncertainty aversion above values of $\gamma = 1/100$, our findings show an increase in out-of-sample loss means. This indicates that the price of robustification is higher than the benefit of avoiding unprofitable models with higher distance to the estimated model.

\begin{rem}[Test hedge]
    Considering the marginal nature of a mere decrease of $0.17\%$ (for $\gamma = 1/100$), we sought to gain a broader understanding of our result by implementing a deep hedge training on a test set generated by the scenarios. Specifically, we utilized each $\sigma$ value, which represents a scenario, to generate a single path. Subsequently, we jointly trained this deep hedge, hereafter referred to as the \textit{test hedge}, on these paths.

    To put this into the context of mathematical formulae, the test hedge represents an agent that knows and optimizes the hedge objective on the combined measure $\Tilde{\bbP}$,
    \begin{align*}
        \myRM^{\Tilde{\bbP}}\Big(\sum_{n = 1}^N \phi^\intercal_{t_{n-1}} \Delta \bfS^{\Tilde{\bbP}}_{t_n} - \bfC_T^{\Tilde{\bbP}}\Big) \longrightarrow \max_\phi !, \quad \Tilde{\bbP} := \frac{1}{M}\bigoplus_{m=1}^M \bbP_m .
    \end{align*}
    Using an optimization over the combined measure is a common economic concept to approach model uncertainty in an ad hoc fashion and is often referred to as Bayesian model averaging, see \cite{HoetingMadiganRafteryVolinksy99} for an introduction, \cite{Cont06} for a critical discussion, as well as \cite{LutkebohmertSchmidtSester21} for an application in the context of deep hedging.

    Note that this hedge is neither optimal with respect to OOSP mean nor OOSP standard deviation, as in general
    \begin{align*}
        \myRM^{\Tilde{\bbP}}\Big(\sum_{n = 1}^N \phi^\intercal_{t_{n-1}} \Delta \bfS^{\Tilde{\bbP}}_{t_n} - \bfC_T^{\Tilde{\bbP}}\Big) &\neq \underbrace{\frac{1}{M} \sum_{m=0}^M \myRM^{\bbP_m}\Big(\sum_{n = 1}^N \phi^\intercal_{t_{n-1}} \Delta \bfS^{\bbP_m}_{t_n} - \bfC_T^{\bbP_m}\Big)}_{=: (\star)} \\
        \myRM^{\Tilde{\bbP}}\Big(\sum_{n = 1}^N \phi^\intercal_{t_{n-1}} \Delta \bfS^{\Tilde{\bbP}}_{t_n} - \bfC_T^{\Tilde{\bbP}}\Big) &\neq \frac{1}{M-1} \sum_{m=0}^M \left(\myRM^{\bbP_m}\Big(\sum_{n = 1}^N \phi^\intercal_{t_{n-1}} \Delta \bfS^{\bbP_m}_{t_n} - \bfC_T^{\bbP_m}\Big) - (\star) \right)^2.
    \end{align*}
    
    Although this synthetic strategy may not necessarily constitute the optimum approach, we observed that incorporating this additional information, which is realistically unavailable to the agent, led to a reduction in out-of-sample mean and variance, as depicted in Table \ref{tab:BlackScholesExample}.
    By considering this enhancement, we can better interpret the improvement achieved. Notably, the robust hedging GAN demonstrates a superior out-of-sample loss performance on average, even surpassing that of the artificial strategy employed.
\end{rem}
\medskip

A more noticeable improvement becomes evident when considering the standard deviation of out-of-sample losses. We see an improvement for all displayed choices of $\gamma$ throughout, with a decrease towards higher uncertainty aversions. The asymptotics have a clear interpretation as well: For $\gamma \to 0$, the values will converge against the ones of the deep hedge, since both coincide in the limit. For $\gamma \to \infty$, yielding a zero strategy, the variance of the out-of-sample loss will be zero. 

We see that high enough uncertainty aversions (for displayed values of $\gamma = 1/75$ and higher), the standard deviation of the out-of-sample loss is even lower than the one of the artificial test hedge. This leaves us with $1/\gamma \in \lbrace 50, 75\rbrace$ where the robust hedging GAN outperforms the test hedge in both, mean and standard deviation of the out-of-sample loss. 

Further details about the out-of-sample performance are visualized for $\gamma = 1/75$ in \Cref{fig:BlackScholesExampleDetail}, where the out-of-sample performances of the individual scenarios are put into perspective. The figure indicates that the robust hedging GAN performs worse in settings with low losses, in this case with smaller (or equal) $\sigma^{\bbP_m}$ for the underlying dynamics than the estimated $\sigma^{\bbQ} = 0.2$. However, considerable improvements are notable for scenarios with high out-of-sample losses, which correspond to under-estimations of the sigma parameter. This aligns with the fact that it is harder to hedge in more volatile markets, a well-known result first presented in \cite{Karoui98}.

\begin{rem}[Ensuring comparability]
    When comparing the deep hedge with its robust hedging GAN extension, it is necessary to ensure that both are trained completely. However, when training with optimizers based on stochastic gradient descent, it is not possible to conclude that training is fully complete. Therefore, we ensured that both components are trained in equal amounts of iterations.

    In detail, we set generate $2^{16}$ paths, and iterate on batches with batch-size in $\lbrace 2^8, 2^{10}, 2^{12}, 2^{14}\rbrace$. We repeat training for five times every batch-size in ascending order, where we regenerate the paths once we looped through them completely. Increasing batch sizes reduces the Monte Carlo error towards the end of training, while the number of iterations is reduced to bound computational efforts.

    Afterwards, we use the pre-trained deep hedge to construct the robust hedging GAN and train it on $1\,000$ epochs of batch-size $2^{16}$. The high batch size is necessary to ensure the accuracy of the penalty term. Eventually, we apply the same training procedure to of $1\,000$ epochs to the pre-trained deep hedge alone to ensure comparability.
\end{rem}

\begin{figure}[t!]
    \centering
    \includegraphics[width=.99\textwidth]{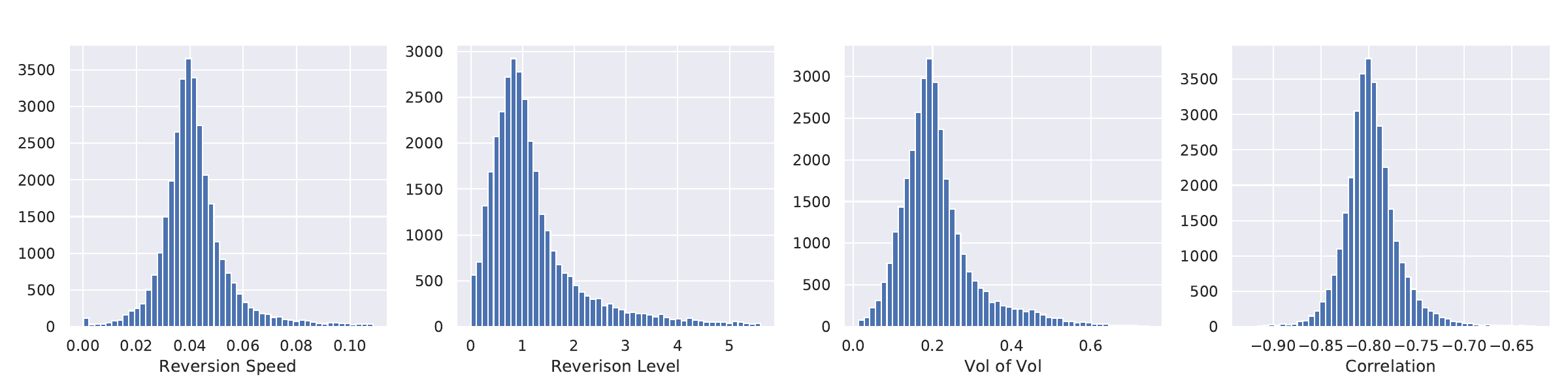}
    \caption{Heston parameters inverse calibrated to vanilla option prices as described in \Cref{exa:CalibrationH} with various data densities.}
    \label{fig:UncertaintyPars}
\end{figure}

\subsection{Market-data induced uncertainty in the Heston model} As a next step, we increase the complexity to the Heston model \cite{Heston93}, where the dynamics are specified by 
\begin{subequations} \label{id:Heston}
\begin{align}
    d\bfS^1_t &= \bfS^1_t  \sqrt{\bfS^2_t} d \bfW^1_t, \quad \forall t \in [0,T] \quad \text{and} \quad \bfS_0^1 \equiv s_0^1, \label{id:Heston1} \\
    d\bfS^2_t &= \kappa^\bbP \left( \beta^\bbP - \bfS^2_t \right) dt + \sqrt{\bfS^2_t}\sigma^\bbP \bfW^2_t, \quad \forall t \in [0,T]  \quad \text{and} \quad \bfS_0^1 \equiv s_0^2\label{id:Heston2}.
\end{align}
The quantities $\kappa^\bbP, \beta^\bbP, \sigma^\bbP \in \R^+$ are positive constants which determine together with the 2-dimensional Brownian motion $\bfW := (\bfW^1, \bfW^2)$, correlated by $\rho^\bbP \in [-1, 1]$, the respective model $\bbP \in \calP$. For simplicity, we set $s^2_0 = \beta^\bbP$. With this, a generator representing a Heston model $\bbP \in \calP$ is fully defined by $\xi := (\kappa^\bbP, \beta^\bbP, \sigma^\bbP, \rho^\bbP) \in \Xi$, where $\Xi := \R_+^3 \times [-1,1]$. 

As the volatility of the process, $\bfS^2$, is not tradable, we introduce another component, namely the volatility swap, that is given by 
\begin{align*}
    \bfS^3_t := \myCond{\bbP}{\int_0^T \bfS^2_s ds}{\sigma((\bfS^1_s, \bfS^2_s)_{s \in [0, t]})},
\end{align*}
since $\bbP$ is a martingale measure. This component can be represented explicitly by
\begin{align}
    \bfS^3_t = \int_0^t \bfS^2_s ds + \frac{\bfS^2_t - \beta^\bbP}{\kappa^\bbP}\left(1 - e^{-\kappa^\bbP(T-t)}\right) + \beta^\bbP(T-t). \label{id:Heston3}
\end{align}
\end{subequations}
In summary, we allow trading on $(\bfS^1, \bfS^3)$, but use the pair $(\bfS^1, \bfS^2)$ as information for the deep hedge and to fit the generator. For more details on this matter, refer to \cite{BuehlerGononTeichmannWood19}.

\begin{exa}[Heston test case]\label{exa:HT}
    We demonstrate the performance of the Heston model analogous to \Cref{exa:BST}, i.e. we use the same time-discretization, European call option to replicate, and hedging objective. We also use the same architecture for the deep hedge. For the test, he estimated model has parameters $\kappa^\bbQ = 1$, $\beta^\bbQ = 0.04$, $\sigma^\bbQ = 0.2$, and $\rho^\bbQ = 0.8$, i.e. $\xi^\bbQ = (\kappa^\bbQ, \beta^\bbQ, \sigma^\bbQ, \rho^\bbQ)$.
    
    The penalty used for the robust hedging GAN in this setting is defined by the $\sigmmd$, which is introduced in Appendix \ref{sec:SigWMetric}. The $\sigmmd$ is applied to the first two components of $\bfS$, i.e. the volatility swap is ignored during its computation. Moreover, the signatures are constructed on paths augmented by time augmentation and a lead-lag transform, see \Cref{rem:Augmentations}, and are of depth $M = 2$. The $\sigmmd$ is scaled by a factor $1/\gamma$ with $\gamma > 0$ when used as penalty.
\end{exa}

\begin{exa}[Heston out-of-sample test data from SPX implied volatilities]\label{exa:CalibrationH}
    A Heston model is usually calibrated to a given implied volatility surface, so an out-of-sample backward test as in \Cref{exa:CalibrationBS} is not directly applicable. 

    Instead, we are mimicking the uncertainty that comes from the fact that the Heston model is calibrated to the volatility surface at initial time and assumes that these are not changing upon maturity. For this, we consider end-of-day market smiles of the SPX index between 1 January 2010 and 18 March 2019 and calibrate a Heston model for each day. To do this efficiently, we used the \textit{deep learning volatility} framework \cite{HorvathMuguruzaTomas19}.

    If $M' \in \N$ is the number of business days in that time span, then the calibrations are expressed by $(\xi_{m})_{m \in \myset{1}{M}} \subseteq \Xi$. We can now consider the change in parameters over $l$ business days, where $l \in \N, l \leq 10$. If we add these differences to our initial choice of parameters, we obtain 
    \begin{align*}
        \xi^{\bbP_{10m + l}} := \xi^\bbQ + (\xi_m - \xi_l)), \quad \forall l \in \myset{1}{10}, \quad \forall m \in \myset{11}{M'}.
    \end{align*}
    The set $(\xi^{\bbP_m})_{m \in \myset{1}{M}}$ with $M = 10M'$ consequently represents how the parameters might change in the next $1-10$ business days, if their relative behaviour has occurred readily in the past.\footnote{Note, that it is not guaranteed that $\xi^{\bbP_m} \in \Xi$ for all $m \in \myset{1}{M}$. In our example, most parameters admitted reasonable values, with only some exceeding the specified domain. As an ad hoc solution, we used the values $\rho^{\bbP_m} \hat{=} \rho^{\bbP_m} \wedge 1$, $\sigma^{\bbP_m} \hat{=} \vert \sigma^{\bbP_m}\vert$ instead for all $m \in \N$.} The distribution of these constructed ``true'' underlying dynamics is visualized in \Cref{fig:UncertaintyPars}. This corresponds to the real life scenario, where an agent calibrated a Heston model to a volatility surface, but refrains from recalibration in the following days. 
\end{exa}

\afterpage{
\begin{table}[p]
    \centering
    \scriptsize
    {\setlength{\tabcolsep}{16pt}
    \begin{tabular}{cccc}
        \toprule
        \textbf{Strategy} &  $1/\gamma$ & \textbf{Mean} & \textbf{Standard Deviation} \\ \midrule
        Deep Hedge       &     & \texttt{0.055450}  &  \texttt{0.011061} \\ \midrule
        \multirow{7}{*}{robust hedging GAN} 
         & ~$\;2\,500$ & \textcolor{C1}{\texttt{0.058885}} &  \textcolor{C2}{\texttt{0.008593}} \\
         & ~$\;5\,000$ & \textcolor{C1}{\texttt{0.055720}} &  \textcolor{C0}{\texttt{0.008648}} \\
         & ~$\;7\,500$ & \textcolor{C1}{\texttt{0.055323}} &  \textcolor{C0}{\texttt{0.009087}} \\
         & $10\,000$   & \textcolor{C0}{\texttt{0.055194}} &  \textcolor{C0}{\texttt{0.009385}} \\
         & $12\,500$   & \textcolor{C0}{\texttt{0.055119}} &  \textcolor{C0}{\texttt{0.009713}} \\
         & $15\,000$   & \textcolor{C0}{\texttt{0.055372}} &  \textcolor{C0}{\texttt{0.009408}} \\
         & $17\,500$   & \textcolor{C2}{\texttt{0.055017}} &  \textcolor{C0}{\texttt{0.009807}} \\
         & $20\,000$   & \textcolor{C0}{\texttt{0.054932}} &  \textcolor{C0}{\texttt{0.009492}} \\
         & $25\,000$   & \textcolor{C0}{\texttt{0.055441}} &  \textcolor{C0}{\texttt{0.009918}} \\ \midrule
        Hedge on Test Set   &     & \texttt{0.056183}  &  \texttt{0.008704} \\ \bottomrule
    \end{tabular}
    }
    \caption{Mean and standard deviation of out-of-sample performances (loss) of the test performed in Example \ref{exa:HT} for various penalty scales.}
    \label{tab:HestonExample}
\end{table}
\begin{figure}[p]
    \centering
    \includegraphics[width=.49\textwidth]{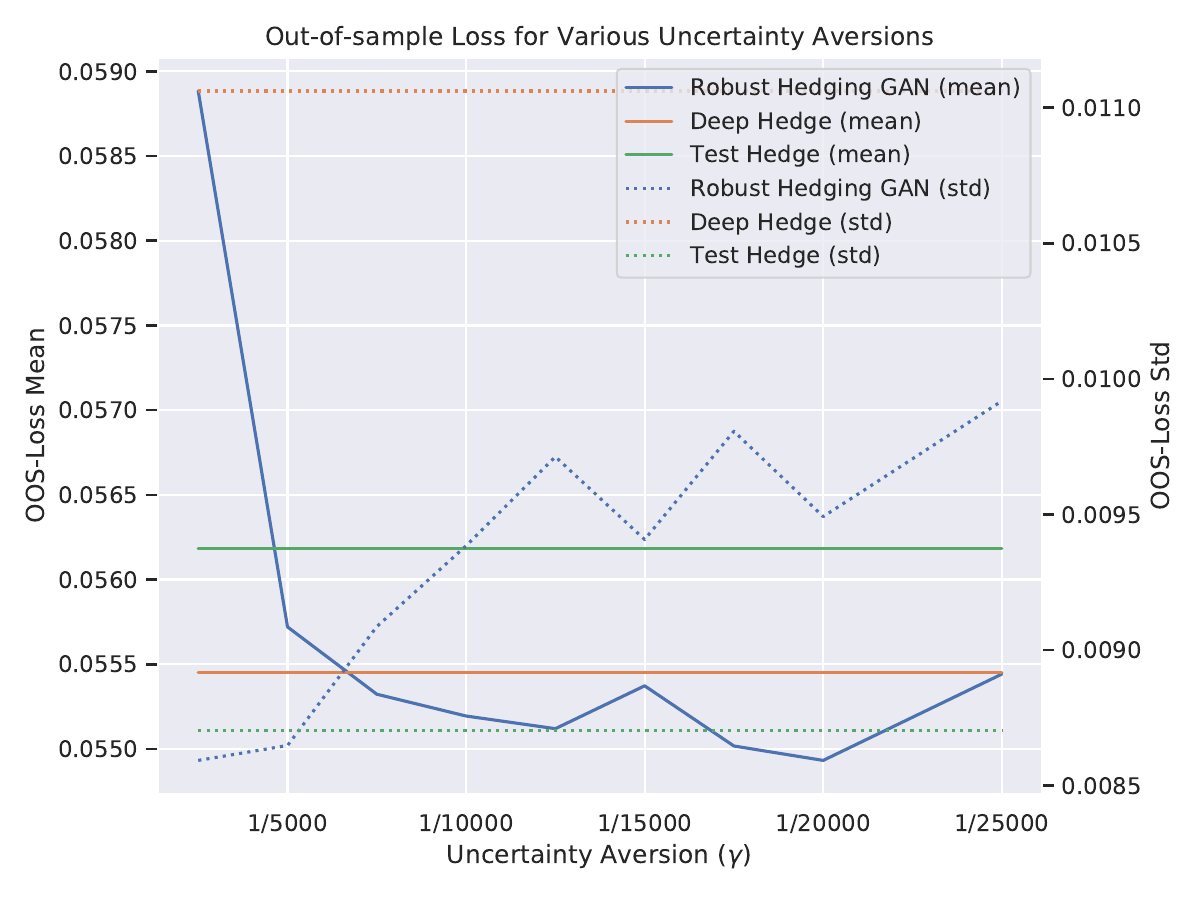}
    \caption{Mean and standard deviation of out-of-sample performances (loss) of the test performed in Example \ref{exa:HT} for various penalty scales.}
    \label{fig:HestonExamplePen}
\end{figure}
\begin{figure}[p]
    \centering
    \includegraphics[width=.99\textwidth]{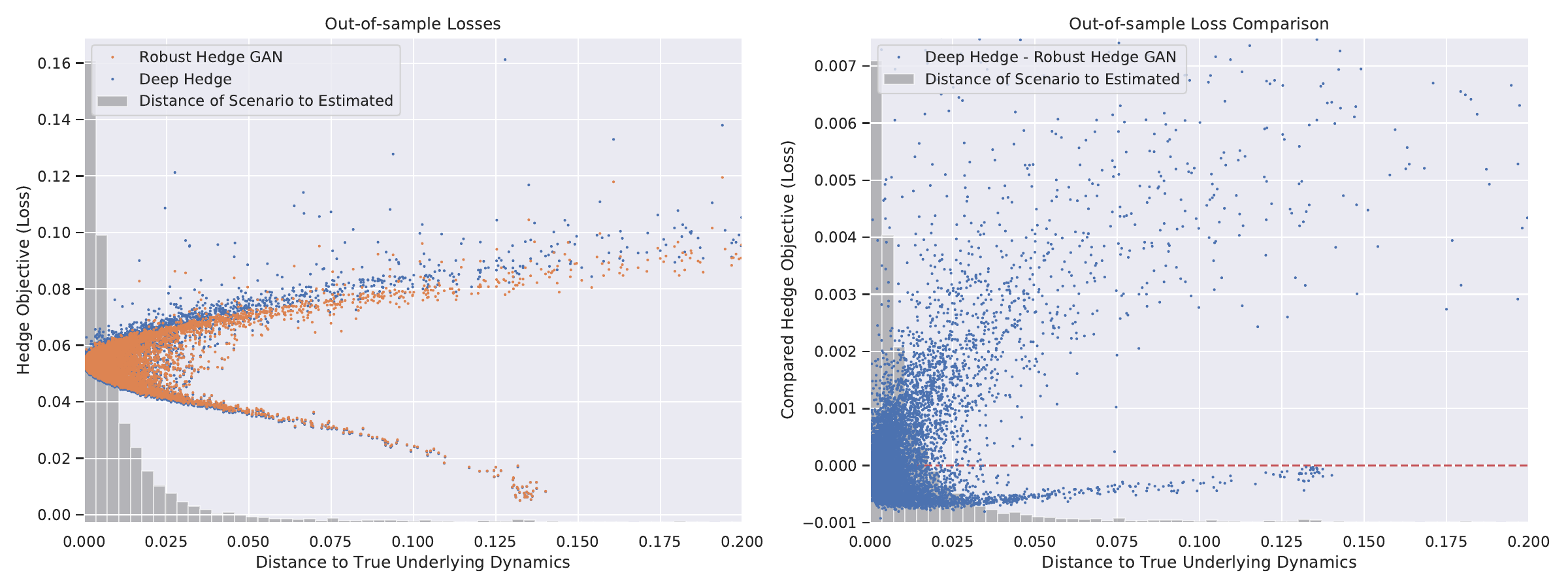}
    \caption{Out-of-sample performances (loss) for individual scenarios of the test performed in Example \ref{exa:HT} with $\gamma := 17\,500$.}
    \label{fig:HestohExampleDetail}
\end{figure}
}

As displayed in \Cref{tab:HestonExample}, the results of \Cref{exa:HT} on data from \Cref{exa:CalibrationH} are supporting our findings from the previous example. The robust hedging GAN has slightly lower expected out-of-sample loss averages, while maintaining a lower variance in these values, both indicating that the robust hedging GAN is favourable in this setting.  

This behaviour is visualized in \Cref{fig:HestonExamplePen}, where we are also able to observe the same implied asymptotic behaviour as in \Cref{exa:BST}: On one hand, the mean of the out-of-sample losses of the robust hedging GAN takes high values for high uncertainty aversions $\gamma$, and converges to the one of the deep hedge for small values of $\gamma$. On the other hand, we can observe the opposite effect for the standard deviation of the robust hedging GAN's out-of-sample losses. While high uncertainty aversion relates here to values close to zero (the strategy of the robust hedging GAN is equivalent to the zero strategy), it increases for low values and limits to the level of the deep hedge.

We also observe a similar interplay with the artificial test hedge as in \Cref{exa:BST}. The robust hedging GAN outperforms it in both, average and standard deviation of the out-of-sample losses, for suitable uncertainty aversions. This also demonstrates the significance of the seemingly minor improvements in the out-of-sample loss average.

In \Cref{fig:HestohExampleDetail} the distribution of the out-of-sample losses in relation to the distance between true underlying dynamics and estimated model are displayed. The relation resembles a parabolic shape, with one branch constituting low loss values and the other in higher loss values. This indicates that a movement of the parameters in one direction benefits the hedge, while the other worsens the performance. As designed, the robust hedging GAN performs better in the high loss regions, what come at the cost of performance in low loss regions.

Altogether, we see that the findings from the simple case of Black-Scholes model uncertainty can be repeated in the more advanced and thus highly intractable setting of Heston model uncertainty.

\subsection{Choice of the uncertainty aversion parameter \texorpdfstring{$\gamma$}{}}

Our analysis does not specifically address the selection of the uncertainty aversion parameter. This choice can be viewed from two perspectives. Firstly, it can be regarded as a decision made by the agent, reflecting their personal preferences. On the other hand, it can also be considered as an optimal choice within the context of a given model and market uncertainty.

To address the optimal choice of the uncertainty aversion parameter within a given model uncertainty framework, we propose a procedure akin to the Heston out-of-sample test above. This approach involves the following steps:
\begin{enumerate}
    \item First, the prevailing model uncertainty is specified as a distribution over the parameter space $\Xi$. This distribution captures the range of potential parameter values that reflect the underlying uncertainty in the market.
    \item Next, a robust hedging GAN is trained using various choices of the uncertainty aversion parameter. The GAN is trained to generate optimal hedging strategies that are resilient to the different levels of uncertainty aversion.
    \item The out-of-sample performance of the trained GAN is then evaluated by comparing the mean and variance of the robust hedging GAN strategies across the different choices of the uncertainty aversion parameter. This comparison allows us to identify a suitable choice of the parameter that yields desirable hedging performance in terms of both average performance and stability.
\end{enumerate}
By following this procedure, market participants can systematically explore and assess the impact of different uncertainty aversion parameter choices on the performance of their hedging strategies. This approach provides a framework for informed decision-making regarding the selection of the uncertainty aversion parameter, taking into account the prevailing model uncertainty and aiming to achieve robust and reliable hedging outcomes in practice.

\subsection{On applying the robust hedging GAN with a NSDE}\label{sec:NSDEApplication} Eventually, we demonstrate the application of the robust hedging GAN with a NSDE as generator. In particular, we use dynamics of type \eqref{id:DynamicsRepeated}, with $d' = 2$. We omit the time component and initialize both nets, approximating $\sigma^\bbP(\cdot)$ and $\mu^\bbP(\cdot)$ respectively, with three hidden layers containing 36 nodes each. To this end, we apply filters to these nets, setting the drift component $\mu^\bbP(\cdot)$ in the first dimension equal to zero and take the absolute value of the diffusion $\sigma^\bbP(\cdot)$ component to ensure non-negative values.

As a first step, we fit the model to the estimated model of Example \ref{exa:HT}, which belongs to the Heston model-class, using the $\sigmmd$-distance. After this, we perform the deep hedge and the robust hedging GAN on both, the NSDE and the Heston model. For this, we use the same derivative, hedge objective, and parameters as in Example \ref{exa:BST} and Example \ref{exa:HT}. However, we only allow trading in the first dimension of the process which is representing the asset price.\footnote{For the sake of simplicity, we also restrict the information for trading to the first component.} Contrary to that, we allow the penalty to operate on both, the asset price process and the volatility process. 

\begin{figure}
    \centering
    \includegraphics[width=.8\textwidth]{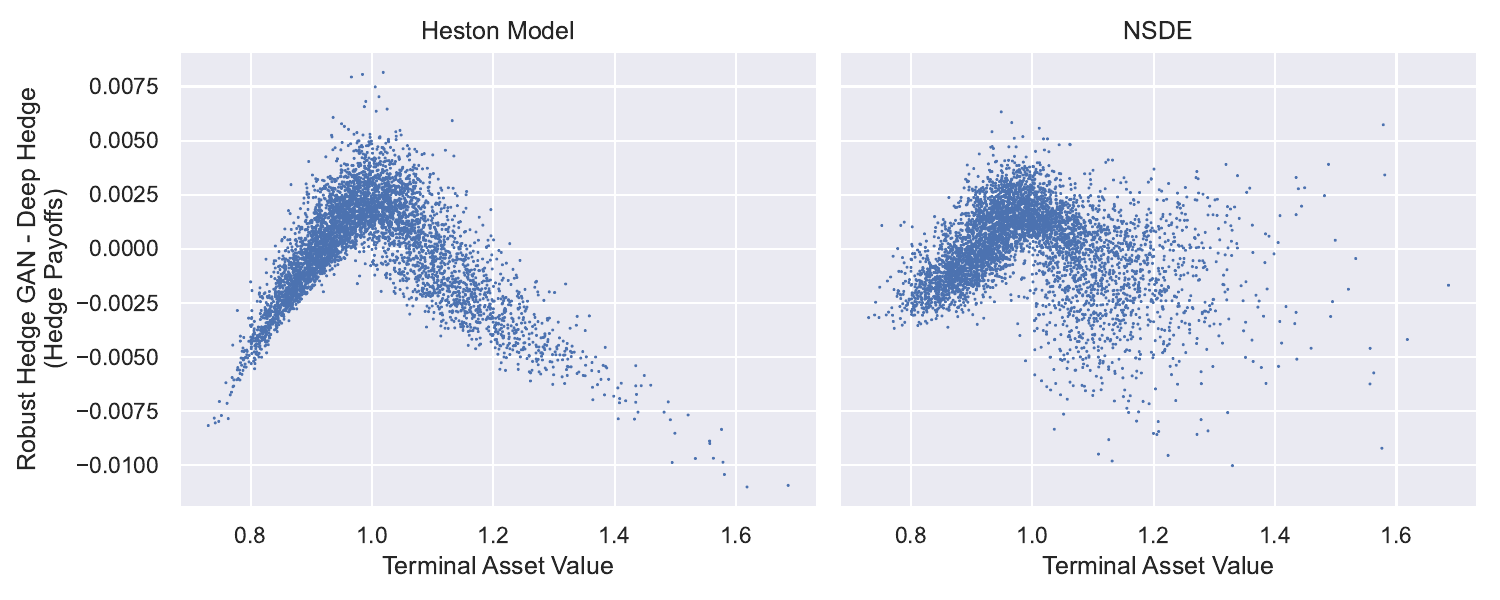}
    \caption{Comparing the robust hedging GAN and the deep hedge for both, an NSDE generator and a Heston generator.}
    \label{fig:NSDEvsHEston.pdf}
\end{figure}

Now, we can observe how the strategy of the robust hedging GAN operates compared to deep hedge in both cases. For this, the difference of the payoffs of the respective strategies is scatter-plotted against the terminal value of the asset price process -- which completely determines the payoff of the derivative -- in Figure \ref{fig:NSDEvsHEston.pdf}. The paths used to generate these payoff differences are sampled from the reference model.

We observe a discernible transformation from the deep hedge strategy to the robust hedging GAN strategy within the Heston model. However, this transformation is not sustained when considering the NSDE generator. The observed trend is consistent with our current understanding, wherein a constrained path generator class accounts for parameter uncertainty, while the high-parametric NSDE approach offers quasi-distributional robustness. This is due to the ability of NSDEs to approximate arbitrary diffusion processes (see \Cref{sec:Generator}), and therefore not being bound to parameter misspecifications as it is the case for the Heston generator class.

\section{Conclusions}

This article introduces an algorithm designed to address problems of the type described in equation \eqref{id:HedgeProblem}. By doing so, we offer a valuable tool for incorporating model uncertainty into the deep hedging framework. Our approach is versatile and can be applied regardless of the chosen data generating process. Additionally, as each component of our algorithm is already utilized in practical contexts, it can be easily adapted to existing functional settings. To illustrate the effectiveness of our approach, we replicated theoretical findings in a simple setting and demonstrated its application in more complex scenarios. Specifically, we successfully enhanced the out-of-sample performance of the deep hedge strategy and compared it to an alternative method that incorporates informed model uncertainty.
\medskip
    

This work is an initial stride towards developing a comprehensive and adaptable framework for addressing the robustification problem in a model-agnostic manner. Building upon these foundations, our future endeavours aim to delve deeper and wider into the exploration of equation \eqref{id:HedgeProblem} and its solutions.

Firstly, we intend to advance theoretical understanding by investigating various aspects, such as the regularity of optimizers and the existence criteria for solutions within different path generator classes. 
Moreover, we envision numerous opportunities for generalizations and extensions within our algorithm. 
Additionally, incorporating extensions of the deep hedge framework itself, as mentioned in Remark \ref{rem:DHE}, can provide new avenues for tackling complex financial scenarios. Ultimately, rigorous testing of our proposed approach on more intricate models, financial products, and real-market data is crucial. Evaluating its performance in non-Markovian settings, exotic options, and other high-complexity contexts will offer valuable insights into its practical applicability. Of particular interest is exploring the advantages of employing more general path generators over traditional equity market models. This shift from parameter uncertainty to quasi-distributional robustness may be a novel solution to the overfitting-problem of high-parametric models.

By pursuing these future directions, we aim to unlock the full potential of our framework as a cutting-edge tool for robustifying financial decision-making, transcending existing limitations.

\bibliographystyle{abbrvnat}
\bibliography{references}


\begin{appendices}
\section{Proofs of Section \ref{sec:Foundation}}\label{sec:Proofsofsection2}

\begin{proof}[Proof of Theorem \ref{thm:Convexity}]
Let $z \in [0,1]$, $\bz := 1-z$ and $\phi, \bphi \in \Phi$. Then,
\begin{align}
    \sup_{\bbP \in \calP} \calL(z\phi + \bz \bphi, \bbP)
    & = \sup_{\bbP \in \calP} \left( \myRM^\bbP\big(z ((\phi \bdot \bfS^{\bbP})_T - \bfC_T^\bbP \big) + \bz ((\bphi \bdot \bfS^{\bbP})_T - \bfC_T^\bbP) \big) - \alpha(\bbP)\right) \notag \\
    & \leq \sup_{\bbP \in \calP} \left( z\myRM^\bbP\big( (\phi \bdot \bfS^{\bbP})_T - \bfC_T^\bbP \big) + \bz \myRM^\bbP\big((\bphi \bdot \bfS^{\bbP})_T - \bfC_T^\bbP  \big) - \alpha(\bbP)\right), \label{id:ExpansionA}
\end{align}
where we applied the convexity of $\myRM^\bbP$ in the last step.  Now, we observe that for functions $f, g: \calP \to \R$ it holds that $\sup_{\bbP \in \calP} f(\bbP) + g(\bbP) \leq \sup_{\bbP \in \calP} f(\bbP) + \sup_{\bbP \in \calP} g(\bbP)$, hence,
\begin{align*}
    \eqref{id:ExpansionA} \leq z\sup_{\bbP \in \calP} \calL(\phi, \bbP) + \bz \sup_{\bbP \in \calP} \calL(\bphi, \bbP), 
\end{align*}
what proves the claim. 
\end{proof}

\begin{proof}[Proof of Theorem \ref{thm:Convexitsecondtheorem}]
Assume that $\bfX_1, \bfX_2 \in \calX$ and let $z \in [0,1]$ with $\bz:= 1- z$. Then,
\begin{align}
    \pi(zX_1 + \bz X_2) &= \inf_{\phi \in \Phi} \left(\sup_{\bbP \in \calP}\left( \myRM^\bbP\big((\phi \bdot \bfS^{\bbP})_T + (z\bfX^\bbP_1 + \bz \bfX^\bbP_2)\big) - \alpha(\bbP)\right)\right) \notag \\
    &= \inf_{\phi, \bphi \in \Phi} \left(\sup_{\bbP \in \calP} \left( \myRM^\bbP\big(z(\phi \bdot \bfS^{\bbP})_T + \bz( \bphi \bdot \bfS^{\bbP})_T + (z\bfX^\bbP_1 + \bz \bfX^\bbP_2)\big) - \alpha(\bbP)\right)\right) \label{id:Expansion1}
\end{align}
due to convexity of $\Phi$. We then apply the convexity of $\myRM^\bbP$ and see that 
\begin{align}
    \eqref{id:Expansion1}
    & \leq \inf_{\phi, \bphi \in \Phi} \left(\sup_{\bbP \in \calP}\left( z\myRM^\bbP\big( (\phi \bdot \bfS^{\bbP})_T  + \bfX^\bbP_1 \big) + \bz \myRM^\bbP\big( (\bphi \bdot \bfS^{\bbP})_T + \bfX^\bbP_2\big) - \alpha(\bbP)\right)\right)  \notag\\
    & = \inf_{\phi, \bphi \in \Phi} \left(\sup_{\bbP \in \calP} \left( z\myRM^\bbP\big( (\phi \bdot \bfS^{\bbP})_T  + \bfX^\bbP_1 \big) - z\alpha(\bbP)+ \bz \myRM^\bbP\big( (\bphi \bdot \bfS^{\bbP})_T + \bfX^\bbP_2\big) - \bz\alpha(\bbP)\right)\right) \label{id:Expansion2}
\end{align}
As before, we use the fact that for functions $f, g: \calP \to \R$ it holds that $\sup_{\bbP \in \calP} f(\bbP) + g(\bbP) \leq \sup_{\bbP \in \calP} f(\bbP) + \sup_{\bbP \in \calP} g(\bbP)$, hence,
\begin{align*}
    \eqref{id:Expansion2} & \leq \inf_{\phi, \bphi \in \Phi}  \left(z \sup_{\bbP \in \calP} \left(\myRM^\bbP\big( (\phi \bdot \bfS^{\bbP})_T  + \bfX^\bbP_1 \big) - \alpha(\bbP) \right)+  \bz \sup_{\bbP \in \calP} \left(\myRM^\bbP\big(( \bphi \bdot \bfS^{\bbP})_T + \bfX^\bbP_2\big) - \alpha(\bbP)\right) \right) \\
    & = z \pi(\bfX_1) + \bz \pi(\bfX_2).
\end{align*}
With that, we have shown that $\pi$ is convex. Furthermore, cash-invariance and monotonicity follows directly from properties of $\myRM^\bbP$.
\end{proof}
\begin{proof}[Proof of Lemma \ref{lem:ThirdLemma}]
    Under the assumption of the Lemma, observe that the 
    \begin{align*}
        p(\bfC_T)
            &= \inf_{\bphi \in \Phi} \sup_{\bbP \in \calP} \Big( \myRM^\bbP\big((\bphi \bdot \bfS^{\bbP})_T - \bfC_T^\bbP \big) - \alpha(\bbP)\Big)     
             -  \inf_{\bphi \in \Phi} \sup_{\bbP \in \calP} \Big( \myRM^\bbP\big((\bphi \bdot \bfS^{\bbP}_T) - \alpha(\bbP)\Big)         \\
            &= \inf_{\bphi \in \Phi} \sup_{\bbP \in \calP} \Big( \myRM^\bbP\big(((\bphi - \phi) \bdot \bfS^{\bbP})_T - p_0 \big) - \alpha(\bbP)\Big)  
             -  \inf_{\bphi \in \Phi} \sup_{\bbP \in \calP} \Big( \myRM^\bbP\big((\bphi \bdot \bfS^{\bbP})_T - \alpha(\bbP)\Big)  \\
            &=  p_0 + \inf_{\bphi \in \Phi} \sup_{\bbP \in \calP} \Big( \myRM^\bbP\big((\bphi \bdot \bfS^{\bbP})_T - \alpha(\bbP)\Big)  
             -  \inf_{\bphi \in \Phi} \sup_{\bbP \in \calP} \Big( \myRM^\bbP\big((\bphi \bdot \bfS^{\bbP})_T - \alpha(\bbP)\Big),
    \end{align*}
    where we used in the second step that $\bphi - \phi \in \Phi$. The above implies $p(\bfC_T) = p_0$ and therefore concludes the proof.
\end{proof}
\begin{proof}[Proof of Theorem \ref{thm:Concavity}]
    Let $\phi \in \Phi$ and $z \in [0,1]$ with $\bz:= 1- z$. Furthermore, fix $\bbP, \bbbP \in \calP$. We can rewrite the loss function as
    \begin{align*}
        \calL(\bbP, \phi) = \inf_{w \in \R} w + \myE{\bbP}{u\big((\phi \bdot \bfS^{\bbP})_T - \bfC_T^\bbP  - w \big)} - \alpha(\bbP) 
    \end{align*}
    and conclude
    \begin{align}
        & \calL(z\bbP + \bz\bbbP, \phi) \notag \\
        &= \inf_{w \in \R} w + z\myE{\bbP}{u\big((\phi \bdot \bfS^{\bbP})_T - \bfC_T^\bbP  - w \big)} + \bz\myE{\bbbP}{u\big((\phi \bdot \bfS^{\bbP})_T - \bfC_T^\bbP  - w \big)} - \alpha(z\bbP + \bz\bbbP) . \label{id:ConcavityExpansion}
    \end{align}
    Next, we observe that for functions $f, g: \R \to \R$ it holds that 
    \begin{align*}
        \inf_{w \in \R} f(w) + g(w) \geq \inf_{w \in \R} f(w) + \inf_{w \in \R} g(w),
    \end{align*}
    so,
    \begin{align*}
    \eqref{id:ConcavityExpansion}
        & \geq z\myRM^\bbP\big((\phi \bdot \bfS^{\bbP})_T - \bfC_T^\bbP\big) + \bz\myRM^{\bbbP}\big((\phi \bdot \bfS^{\bbP})_T - \bfC_T^\bbP\big) - \alpha(z\bbP + \bz\bbbP) \\
        & \geq z \calL(\phi, \bbP) + \bz \calL(\phi, \bbbP),
    \end{align*}
    where we employed convexity of $\alpha$ in the last step. This proves the assertion.
\end{proof}

\section{Rough path theory and distances of stochastic processes}\label{sec:Distances}

The objective of this section is to revisit the concepts of path and expected signatures of a stochastic process, which characterize it and its path law respectively, to establish a distance on stochastic processes. While our framework is capable of employing various notions of distances, the following concepts first established in rough path theory deserve particular mention, since they are model-independent and have proven to be effective in a range of applications, see for instance \cite{KidgerBonnierPerezArribasSalviLyons19, NiSzpruchWieseShujian20, ArribasSalviSzpruch20}.   

In this work, we restrict ourselves to the class of continuous paths on a time interval $\TI \subseteq \R$ with values in $\R^d$ and finite $p$-variation, i.e. $\calC^p_0(\TI, \R^d)$. For a detailed discussion of the signature feature set for parametrized paths, refer to \cite{Lyons14, ChevyrevKormilitzin16, LevinLyonsNi13}.

\subsection{The signature of a path}
Fix by $T((\R^d)) := \oplus_{k = 0}^\infty(\R^d)^{\otimes k}$ a tensor algebra, the space the signature of a $\R^d$-valued path takes values in. Then, the signature of a path is defined as follows.\footnote{For more background on this, refer to \cite{ChevyrevKormilitzin16} or the appendix of \cite{NiSzpruchSabatevidalesWieseXiaoShujian20} and the references mentioned therein.}
\begin{defn}[Signature]
    Let $X \in \calC_0^1(\TI, \R^d)$ such that the integrals in \eqref{id:Signature} are well-defined. The \emph{signature of the path} $X$ is defined as $\sig(X_J) := (1, X_J^1, \ldots, X^k_J, \ldots) \in T((\R^d))$, where
    \begin{align}
        X_J^k = \underset{{(t_1, \ldots, t_k) \in J^k},\, {t_{1} < \ldots <  t_k}}{\int \cdots \int} dX_{t_1} \otimes \cdots \otimes dX_{t_k},\quad k \in \N. \label{id:Signature}
    \end{align}
    Moreover, let us denote the \emph{truncated signature of the path} $X$ at degree $M \in \N$ by $\sig_M(X_J) := (1, X^1_J, \ldots, X^M_J)$.
\end{defn}

The first feature of the signature map that we want to highlight is \emph{uniqueness}. In particular, a signature can determine the path up to tree-like equivalence, see \cite{HamblyLyons10, BoedihardjoGeng15}. However, since the path signature is invariant with respect to time re-parametrizations and its starting value, it is necessary to augment the observed path before applying the feature map.

\begin{rem}[Path augmentations]\label{rem:Augmentations}
    Common path augmentations that have been suggested in \cite{MorrillFermanianKidgerLyons20} include
    \begin{enumerate*}[label=(\alph*)]
        \item \label{item:AugTime} the time augmentation, 
        \item \label{item:AugVisi} the visibility transform, and 
        \item \label{item:AugLeLa} the lead-lag transform.
    \end{enumerate*}
    These augmentations encode the time parametrization, the starting point, and the lagged process into the path by adding further dimensions. In detail, a path of dimension $d$ results into a path taking values in $\R^{2d + 1}$ after applying augmentations \ref{item:AugTime}, \ref{item:AugVisi}, and \ref{item:AugLeLa}.\footnote{For the sake of simplicity, we abuse the notation and interpret $\calC^1_0(\TI, \R^d)$ either as the space of paths or augmented paths, depending on the context.}
\end{rem}

Another feature of the signature map is \emph{universality}, that means that every continuous functional on the path space $\calC^1_0(\TI, \R^d)$ can be approximated arbitrarily well by linear functionals of truncated signatures when the degree of the signature is high enough. This is summarized in the subsequent theorem.
\begin{thm}[Universality of Signatures \cite{LevinLyonsNi13}]\label{thm:Universality}
    For any continuous functional $f: K \to \R$, where $K \subseteq \calC^1_0(\TI, \R^d)$ is a compact set, and any $\epsilon > 0$, there is a functional $L \in T((\R^d))^*$ such that 
    \begin{align*}
        \sup_{X \in K} \vert f(X) - L(\sig(X)) \vert < \epsilon.
    \end{align*}
\end{thm}

\begin{rem}[Implementation]
    Since data of stochastic processes is usually provided in a discretized form, we use linear interpolation between the values to elevate the data to continuous paths. To compute the signatures, the open source library signatory \cite{Signatory} which operates in PyTorch \cite{PyTorch} is used.
\end{rem}

\subsection{The expected signature of a stochastic process}

Let $\bfX$ be a stochastic process defined on a probability space $(\Omega, \calF, \bbP)$ with paths in $\calC^p_0(\TI, \R^d)$, such that \eqref{id:Signature} is well-defined almost surely. Then, by path-wise applying the signature map, we obtain the expected signature $\myE{\bbP}{\sig(\bfX)}$. Of course, the expected signature is only well-defined if $\sig(\bfX)$ has finite expectation w.r.t. $\bbP$. The following result states that the expected signature characterizes the law of the stochastic process.

\begin{thm}[\cite{ChevyrevLyons16}]
    If $\bfX$ and $\bfY$ are two $\calC_0^1(\TI, \R^d)$-valued random variables, $\myE{\bbP}{\sig(\bfX)} = \myE{\bbP}{\sig(\bfY)}$, and $\myE{\bbP}{\sig(\bfX)}$ has a finite radius of convergence, then $\bfX \overset{d}{=} \bfY$.
\end{thm}

To put this another way, under the regularity condition, the distribution of some process $\bfX$ under $\bbP$ on the path space is characterized by $\myE{\bfX \sim \bbP}{\sig(\bfX)}$. This means, the expected signature of a random process can be viewed as an analogy of the moment generating function of a $d$-dimensional random variable. In \cite{LyonsNi15} it is for instance shown that the expected Stratonovich signature of a Brownian motion determines its law. For a more detailed discussion on the characterization of stochastic processes by moments of signatures, see \cite{ChevyrevOberhauser18}.

\subsection{Signature Wasserstein-1 metric (\texorpdfstring{$\sigw$}{Sig-W1}) and Signature-MMD (\texorpdfstring{$\sigmmd$}{Sig-MMD})}\label{sec:SigWMetric}
As a next step, we will briefly discuss the metrics which are used to construct the penalty term $\alpha$ in \eqref{id:HedgeProblem} that are based on the signature framework. We begin by defining the Sig-Wasserstein distance. 
Note that in certain circumstances the following distances, introduced in recent literature are equivalent to one-another, however their numerical approximation and computational efficiency may have some discrepancies in applications. We therefore include a reminder and analysis with both metrics here and postpone the discussion of consolidating our findings to the identification of the ideal metric to future work.

\begin{defn}[The Signature Wasserstein-1 metric]
    Assume $\bbP, \bbQ$ are two measures on the path space $\calC^1_0(\TI, \R^d)$ and $\bfX$ is a stochastic process taking values in this space, then the \emph{signature Wasserstein-1 metric on path space} ($\sigw$), is defined as
    \begin{align*}
        \sigw (\bbP, \bbQ) = \sup_{\Vert f \Vert_{\mathrm{Lip}} \leq 1} \myE{\bfX\sim\bbP}{f(\sig(\bfX))} - \myE{\bfX\sim\bbQ}{f(\sig(\bfX))},
    \end{align*}
    where the supremum is over all the $1$-Lipschitz functions $f: \calC^1_0(\TI, \R^d) \to \R$ with Lipschitz norm smaller than 1, i.e. for all $x_1, x_2  \in \calC^1_0(\TI, \R^d)$ it holds that $\vert f(x_1)-f(x_2)\vert \leq \vert x_1-x_2\vert$.
\end{defn}

This distance is based on the fact that signatures are representing stochastic processes in a random variable-like manner and the fact that the Wasserstein-metric provides a way to measure the distance between vector valued random variables. This metric goes back to \cite{Vasershtein69}. While our framework is capable of employing a penalty that is defined via a maximization problem (see Section \ref{sec:RobustHedgeGAN}), we follow the steps presented in \cite{NiSzpruchWieseShujian20, NiSzpruchSabatevidalesWieseXiaoShujian20} to find and approximation of this metric and incorporate the analytic solution thereof. We begin with an application of \Cref{thm:Universality},
\begin{align}
    \sigw (\bbP, \bbQ) = \sup_{\overset{\Vert L \Vert \leq 1}{L\textit{ is linear}}} \myE{\bfX\sim\bbP}{L(\sig(\bfX))} - \myE{\bfX\sim\bbQ}{L(\sig(\bfX))}, \label{id:SigW1metric}
\end{align}
where the norm on $L$ is chosen as the $l_2$-norm of the linear coefficients in $L$. Then we approximate $\sigw (\bbP, \bbQ)$ by $\sigw^{(M)}(\bbP, \bbQ)$, where we truncate the signatures in \eqref{id:SigW1metric} at level $M \in \N$. Since functionals on finite dimensional spaces can be represented by a vector multiplication, the optimization problem admits the analytic solution
\begin{align}
    \sigw^{(M)} (\bbP, \bbQ) = \left\vert \myE{\bfX\sim\bbP}{\sig_M(\bfX)} - \myE{\bfX\sim\bbQ}{\sig_M(\bfX)}\right\vert. \label{id:SigW1MetricAnalytic}
\end{align}
where $\vert \cdot \vert$ is the $l_2$ norm on the truncated tensor algebra $T^M((\R^d))$ and $M \in \N$ is the truncation level. For more details on this derivation, see \cite{BalajiSajediKalibhatDingStoegerSoltanolkotabiFeizi21, NiSzpruchWieseShujian20}.

Eventually, note that the above derivation leads to the square of \eqref{id:SigW1MetricAnalytic} for the \textit{signature maximum mean discrepancy} ($\sigmmd$) as defined in \cite{ChevyrevOberhauser18} if one chooses the signature with truncation level $M$ as feature map, see \cite{NiSzpruchWieseShujian20}.
    
\begin{rem}[Kernel trick and PDE representation]\label{rem:ME}
    The approximation \eqref{id:SigW1MetricAnalytic} relies on the accuracy of \Cref{thm:Universality} and in practice, where low values of $M$ are used, this may lead to larger error terms. Moreover, to obtain sufficient precision for Monte-Carlo estimates of the expectations in \eqref{id:SigW1MetricAnalytic}, high sample sizes may be necessary to avoid significant bias.\footnote{Our findings suggest that low sample sizes tend to under-estimate the volatility of the underlying process.}
    
    There are efforts to circumvent this approximation whilst maintaining a computable metric: As it is presented in \cite{ChevyrevOberhauser18}, the $\sigmmd$ has a representation using the signature kernel, for which \citeauthor{SalviCassFosterLyonsYang21} in \cite{SalviCassFosterLyonsYang21} show that it is possible to compute as the solution of a Gorsuat-PDE. This leads to potential speed-ups, as it is indicated by the benchmarking performed in \cite{HorvathIssaLemercierSalvi23}. 
\end{rem}

\section{Out-of-sample Testing} \label{sec:Test}

In this section, we investigate an out-of-sample test to evaluate a hedging strategy in an uncertainty setting. A straightforward approach is to fix some true underlying dynamics and perform the hedge-algorithm, as an agent would, on small data environments thereof and evaluate the strategies on the true dynamics.

To make this precise, stipulate that $\calG_{\xi^*}$ is a generator for the true underlying dynamics for a fixed parameter $\xi^* \in \Xi$. This generator is not available to the actors, i.e. its data is not used to train components of the deep hedge. Moreover, we let $\dist$ be an arbitrary \emph{calibration metric}, that is a function
\begin{align*}
    \dist: 
    \begin{array}{ccl}
    \R^{\vert I \vert \times N + 1 \times d'} \times \Xi & \longrightarrow & \R \\
    \left((S^I_{t_n})_{n \in \lbrace 0, \ldots, N \rbrace}, \xi \right) & \longmapsto & \dist\left((S^I_{t_n})_{n \in \lbrace 0, \ldots, N \rbrace}, \xi \right),
    \end{array}
\end{align*}
that determines the distance of some set of paths $(S^I_{t_n})_{n \in \lbrace 0, \ldots, N \rbrace}$ indexed by some index set $I \subseteq \N$ to the generator's parameter $\xi$. The reader may for instance think of a Black-Scholes type generator (see \Cref{exa:BST}) and $\dist$ the absolute distance of the (average) realized volatility on $(S^I_{t_n})_{n \in \lbrace 0, \ldots, N \rbrace}$ to the generator's volatility parameter $\sigma = \xi$.

The test can then be described with the following three steps:
\begin{enumerate}
    \item \text{Scenario generation.}
    \begin{enumerate}
        \item We generate $M \in \N$ batches of Brownian increments,  $(\Delta W^{I, m}_{t_n})_{n \in \lbrace 1, \ldots, N \rbrace}$ where $m \in \lbrace 1, ..., M \rbrace$ refers to the batch which is then indexed by some index set $I \subseteq \N$. We will refer to each of the $M$ batches as a \emph{scenario}. 
        \item Next, the generator is applied to the batches of increments to obtain batches of paths
        \begin{align*}
            (S^{I, m}_{t_n})_{n \in \lbrace 1, ..., N \rbrace} := \left\lbrace\calG_{\xi^*}\left((\Delta W^{m, i}_{t_n})_{n \in \lbrace 1, \ldots, N \rbrace}\right) \right\rbrace_{i \in I}
        \end{align*}
        for every $m \in \myset{1}{M}$. We refer to a batch as a scenario.
        \item \label{item:UncertaintySource} For every scenario $m \in \myset{1}{M}$, we fit a generator $\calG_{\cdot}$ to the corresponding batch of paths. This means to find for every  a parameter $\xi_m \in \Xi$ s.t.
        \begin{align*}
            \xi_m := \argmin_{\xi \in \Xi} \dist \left((S^{I, m}_{t_n})_{n \in \lbrace 0, \ldots, N \rbrace}, \xi \right).
        \end{align*}
    \end{enumerate}
    \item \text{Training.}
    \begin{enumerate}
        \item For each scenario $m \in \myset{1}{M}$, we train a deep hedge on the respective generator $\calG_{\xi_m}$. 
        \item Likewise, we use the generator $\calG_{\xi^m}$ for each scenario $m \in \myset{1}{M}$ to construct the robust hedging GAN and train it adversarially.
    \end{enumerate}
    \item \text{Evaluation.}
    We can now evaluate the out-of-sample performance (OOSP) (implemented as loss) of both approaches. For this, let $\lbrace \phi^{(m)} \rbrace_{m \in \lbrace 1, \ldots, M \rbrace}$ correspond to the strategies arising from the $M$ batches. If $\bbP^*$ is the measure that matches the dynamics implied by the reference generator $\calG_\xi$, the out-of-sample performance corresponds to
    \begin{align}
        \bigg\lbrace - \myRM^{\bbP^*}\Big(\sum_{n = 1}^N {\phi^{(m)}_{t_{n-1}}}^\intercal \Delta \bfS^{\bbP^*}_{t_n} - \bfC_T\Big)\bigg\rbrace_{m \in \lbrace 1, \ldots, M \rbrace}. \label{id:OOSP}
    \end{align}
\end{enumerate}
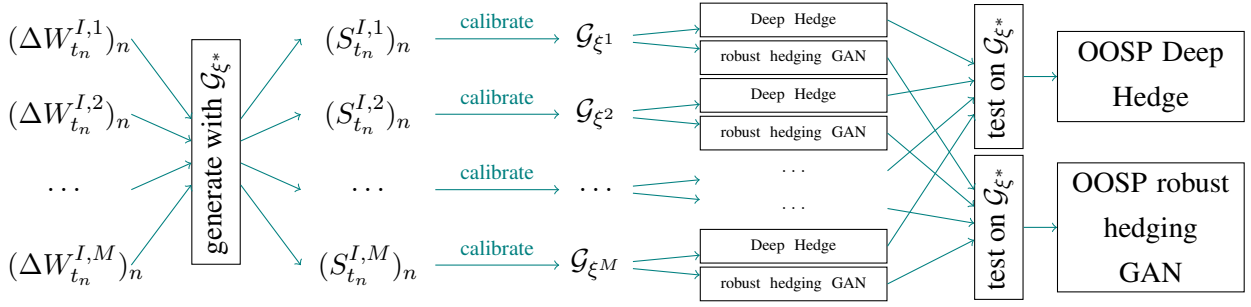
\begin{figure}
    \centering
    \small
        \begin{tikzpicture}
            \node[align=center, text width=1.5cm] at (0, 3) (bi1) {$(\Delta W^{I, 1}_{t_n})_{n}$};
            \node[align=center, text width=1.5cm] at (0, 2) (bi2) {$(\Delta W^{I, 2}_{t_n})_{n}$};
            \node[align=center, text width=1.5cm] at (0, 0) (bi3) {$(\Delta W^{I, M}_{t_n})_{n}$};
            \node[align=center, text width=1.5cm] at (0, 1) (bid) {$\ldots$};

            \node[align=center, text width=1.5cm] at (4, 3) (paths1) {$(S^{I, 1}_{t_n})_{n}$};
            \node[align=center, text width=1.5cm] at (4, 2) (paths2) {$(S^{I, 2}_{t_n})_{n}$};
            \node[align=center, text width=1.5cm] at (4, 0) (paths3) {$(S^{I, M}_{t_n})_{n}$};
            \node[align=center, text width=1.5cm] at (4, 1) (pathsd) {$\ldots$};
            
            \node[rotate=90, draw, fill=white] at (2, 1.5) (refgen) {generate with $\calG_{\xi^*}$};
            
            \draw[->, teal] (bi1.east) -- (refgen);
            \draw[->, teal] (bi2.east) -- (refgen);
            \draw[->, teal] (bi3.east) -- (refgen);
            \draw[->, teal] (bid.east) -- (refgen);
            \draw[->, teal] (refgen) -- (paths1.west);
            \draw[->, teal] (refgen) -- (paths2.west);
            \draw[->, teal] (refgen) -- (paths3.west);
            \draw[->, teal] (refgen) -- (pathsd.west);
            
            \node[align=center, text width=.7cm] at (7, 3) (gen1) {$\calG_{\xi^1}$};
            \node[align=center, text width=.7cm] at (7, 2) (gen2) {$\calG_{\xi^2}$};
            \node[align=center, text width=.7cm] at (7, 0) (gen3) {$\calG_{\xi^M}$};
            \node[align=center, text width=.7cm] at (7, 1) (gend) {$\ldots$};
            \draw[->, teal] (paths1) -- node[above, midway] {\scriptsize calibrate} (gen1);
            \draw[->, teal] (paths2) -- node[above, midway] {\scriptsize calibrate} (gen2);
            \draw[->, teal] (paths3) -- node[above, midway] {\scriptsize calibrate} (gen3);
            \draw[->, teal] (pathsd) -- node[above, midway] {\scriptsize calibrate} (gend);

            \node[align=center, text width=2.2cm, draw] at (9.6, 3.25) (dh1) {\tiny Deep Hedge};
            \node[align=center, text width=2.2cm, draw] at (9.6, 2.25) (dh2) {\tiny Deep Hedge};
            \node[align=center, text width=2.2cm, draw] at (9.6, 0.25) (dh3) {\tiny Deep Hedge};
            \node[align=center, text width=2.2cm,     ] at (9.6, 1.25) (dhd) {\tiny $\ldots$};
            \node[align=center, text width=2.2cm, draw] at (9.6, 2.75) (rh1) {\tiny robust hedging GAN};
            \node[align=center, text width=2.2cm, draw] at (9.6, 1.75) (rh2) {\tiny robust hedging GAN};
            \node[align=center, text width=2.2cm, draw] at (9.6, -.25) (rh3) {\tiny robust hedging GAN};
            \node[align=center, text width=2.2cm,     ] at (9.6, 0.75) (rhd) {\tiny $\ldots$};
            \draw[->, teal] (gen1) -- (dh1);
            \draw[->, teal] (gen2) -- (dh2);
            \draw[->, teal] (gen3) -- (dh3);
            \draw[->, teal] (gend) -- (dhd);
            \draw[->, teal] (gen1) -- (rh1);
            \draw[->, teal] (gen2) -- (rh2);
            \draw[->, teal] (gen3) -- (rh3);
            \draw[->, teal] (gend) -- (rhd);
            
            \node[rotate=90, draw, fill=white] at (12.3, 2.5) (testdh) {test on $\calG_{\xi^*}$};
            \node[rotate=90, draw, fill=white] at (12.3, .5) (testrh) {test on $\calG_{\xi^*}$};
            \draw[->, teal] (dh1.east) -- (testdh);
            \draw[->, teal] (dh2.east) -- (testdh);
            \draw[->, teal] (dh3.east) -- (testdh);
            \draw[->, teal] (dhd.east) -- (testdh);
            \draw[->, teal] (rh1.east) -- (testrh);
            \draw[->, teal] (rh2.east) -- (testrh);
            \draw[->, teal] (rh3.east) -- (testrh);
            \draw[->, teal] (rhd.east) -- (testrh);
            
            \node[align=center, text width=2.2cm, draw] at (14.3, 2.5) (dho) {OOSP Deep Hedge};
            \node[align=center, text width=2.2cm, draw] at (14.3, 0.5) (rho) {OOSP robust hedging GAN};

            \draw[->, teal] (testdh) -- (dho);
            \draw[->, teal] (testrh) -- (rho);
		\end{tikzpicture}
    \caption{Structure of the out-of-sample performance test on scenarios $\myset{1}{M}$.}
    \label{fig:TestA}
\end{figure}
The test structure is visualized in \Cref{fig:TestA}. It is clear, that the set size $\vert I \vert$ dictates the level of model uncertainty that the scenarios inherit, while $M$ indicates the precision when statistically evaluating \eqref{id:OOSP}.

\paragraph{Backward test} The above introduced test comes with a high computational demand, since it is necessary to retrain deep hedge and robust hedging GAN for each scenario. Instead, we suggest calibrating different generators to one model, where in every case the former may be the measure indicating the true underlying for the latter. 

To make this precise, assume that we estimated a measure $\bbQ$ under which we know the dynamics of $(\bfS^{\bbQ}_{t_n})_{n \in \lbrace 0, \ldots, N \rbrace}$. These are expressed by the generator $\calG_{\xi^*}$ for some parameter ${\xi^*} \in \Xi$. With that, the test reads as follows:
\begin{enumerate}
    \item Scenario generation.
    \begin{enumerate}
        \item As before, we generate $M \in \N$ batches of Brownian increments,  $(\Delta W^{I, m}_{t_n})_{n \in \lbrace 1, \ldots, N \rbrace}$. Hereby, $m \in \lbrace 1, ..., M \rbrace$ refers to the batch which is then indexed by some index set $I \subseteq \N$. We will refer to each of the $M$ batches as a \emph{scenario}. 
        \item For every $m \in \myset{1}{M}$ we now find parameters $\xi^m$ such that $\calG_{\xi_m}$ represents the true underlying distribution in this scenario
        \begin{align*}
            \xi_m := \argmin_{\xi \in \Xi} \dist \left(\left\lbrace\calG_{\xi}\left((\Delta W^{m, i}_{t_n})_{n \in \lbrace 1, \ldots, N \rbrace}\right) \right\rbrace_{i \in I}, \xi^*\right).
        \end{align*}
    \end{enumerate}
    \item Training.
    \begin{enumerate}
        \item We train a deep hedge on the generator representing the estimated model $\calG_{\xi^*}$. 
        \item Likewise, we use the estimated generator $\calG_{\xi^*}$ to construct the robust hedging GAN and train it adversarially.
    \end{enumerate}
    \item Evaluation.
    We can now evaluate the out-of-sample performance (OOSP) (implemented as loss) of both approaches. For this, let $\lbrace \phi \rbrace_{m \in \lbrace 1, \ldots, M \rbrace}$ correspond to a strategy from one of the both procedures. If for every $m \in \myset{1}{M}$, $\bbP^m$ is the measure that matches the dynamics implied by the generator $\calG_{\xi_m}$, the out-of-sample performance corresponds to
    \begin{align}
        \bigg\lbrace - \myRM^{\bbP^m}\Big(\sum_{n = 1}^N {\phi_{t_{n-1}}}^\intercal \Delta \bfS^{\bbP^m}_{t_n} - \bfC_T\Big)\bigg\rbrace_{m \in \lbrace 1, \ldots, M \rbrace}. \label{id:OOSP2}
    \end{align}
\end{enumerate}
\begin{figure}
    \centering
    \small
        \begin{tikzpicture}
            \node[align=center, text width=1.5cm] at (0, 3) (bi1) {$(\Delta W^{I, 1}_{t_n})_{n}$};
            \node[align=center, text width=1.5cm] at (0, 2) (bi2) {$(\Delta W^{I, 2}_{t_n})_{n}$};
            \node[align=center, text width=1.5cm] at (0, 0) (bi3) {$(\Delta W^{I, M}_{t_n})_{n}$};
            \node[align=center, text width=1.5cm] at (0, 1) (bid) {$\ldots$};

            \node[align=center, text width=1.5cm] at (4, 3) (paths1) {$(S^{I, 1}_{t_n})_{n}$};
            \node[align=center, text width=1.5cm] at (4, 2) (paths2) {$(S^{I, 2}_{t_n})_{n}$};
            \node[align=center, text width=1.5cm] at (4, 0) (paths3) {$(S^{I, M}_{t_n})_{n}$};
            \node[align=center, text width=1.5cm] at (4, 1) (pathsd) {$\ldots$};
            
            \node[align=center, text width=.7cm] at (2, 3) (gen1) {$\calG_{\xi^1}$};
            \node[align=center, text width=.7cm] at (2, 2) (gen2) {$\calG_{\xi^2}$};
            \node[align=center, text width=.7cm] at (2, 0) (gen3) {$\calG_{\xi^M}$};
            \node[align=center, text width=.7cm] at (2, 1) (gend) {$\ldots$};
            
            
            \draw[->, teal] (bi1.east) -- (gen1);
            \draw[->, teal] (bi2.east) -- (gen2);
            \draw[->, teal] (bi3.east) -- (gen3);
            \draw[->, teal] (bid.east) -- (gend);
            \draw[->, teal] (gen1) -- (paths1.west);
            \draw[->, teal] (gen2) -- (paths2.west);
            \draw[->, teal] (gen3) -- (paths3.west);
            \draw[->, teal] (gend) -- (pathsd.west);
            
            \node[align=center, text width=.7cm] at (6, 1.5) (estgen) {$\calG_{\xi^*}$};
            
            \draw[<-, purple] (gen1.south) -- ($(gen1.south)+(0, -.3)$) -- node[above, midway] {\tiny inverse calibrate} ($(gen1.south)+(3, -.3)$) -- (estgen); %
            \draw[<-, purple] (gen2.south) -- ($(gen2.south)+(0, -.3)$) -- node[above, midway] {\tiny inverse calibrate} ($(gen2.south)+(3, -.3)$) -- (estgen); %
            \draw[<-, purple] (gen3.south) -- ($(gen3.south)+(0, -.3)$) -- node[above, midway] {\tiny inverse calibrate} ($(gen3.south)+(3, -.3)$) -- (estgen); %
            \draw[<-, purple] (gend.south) -- ($(gend.south)+(0, -.3)$) -- node[above, midway] {\tiny inverse calibrate} ($(gend.south)+(3, -.3)$) -- (estgen); %

            \draw[->, teal, dashed] (paths1.east) -- (estgen);
            \draw[->, teal, dashed] (paths2.east) -- (estgen);
            \draw[->, teal, dashed] (paths3.east) -- (estgen);
            \draw[->, teal, dashed] (pathsd.east) -- (estgen);
            
            \node[align=center, text width=2.2cm, draw] at (8, 2.5) (dh) {\tiny Deep Hedge};
            \node[align=center, text width=2.2cm, draw] at (8, .5) (rh) {\tiny robust hedging GAN};
            \draw[->, teal] (estgen) -- (dh.west);
            \draw[->, teal] (estgen) -- (rh.west);
            
            \node[align=center, text width=1.75cm, draw] at (11.2, 3) (test1) {\footnotesize test on $\calG_{\xi^1}$};
            \node[align=center, text width=1.75cm, draw] at (11.2, 2) (test2) {\footnotesize test on $\calG_{\xi^2}$};
            \node[align=center, text width=1.75cm, draw] at (11.2, 0) (test3) {\footnotesize test on $\calG_{\xi^M}$};
            \node[align=center, text width=1.75cm,     ] at (11.2, 1) (testd) {\footnotesize $\ldots$};
            
            \draw[->, teal] (dh.east) -- (test1.west);
            \draw[->, teal] (dh.east) -- (test2.west);
            \draw[->, teal] (dh.east) -- (test3.west);
            \draw[->, teal] (dh.east) -- (testd.west);
            \draw[->, teal] (rh.east) -- (test1.west);
            \draw[->, teal] (rh.east) -- (test2.west);
            \draw[->, teal] (rh.east) -- (test3.west);
            \draw[->, teal] (rh.east) -- (testd.west);
            
            \node[align=center, text width=2.2cm, draw] at (14.3, 2.5) (dho) {OOSP Deep Hedge};
            \node[align=center, text width=2.2cm, draw] at (14.3, 0.5) (rho) {OOSP robust hedging GAN};

            \draw[->, teal] (test1.east) -- (dho.west);
            \draw[->, teal] (test2.east) -- (dho.west);
            \draw[->, teal] (test3.east) -- (dho.west);
            \draw[->, teal] (testd.east) -- (dho.west);
            \draw[->, teal] (test1.east) -- (rho.west);
            \draw[->, teal] (test2.east) -- (rho.west);
            \draw[->, teal] (test3.east) -- (rho.west);
            \draw[->, teal] (testd.east) -- (rho.west);
            
		\end{tikzpicture}
    \caption{Structure of the inverse out-of-sample performance test on scenarios $\myset{1}{M}$.}
    \label{fig:TestB}
\end{figure}
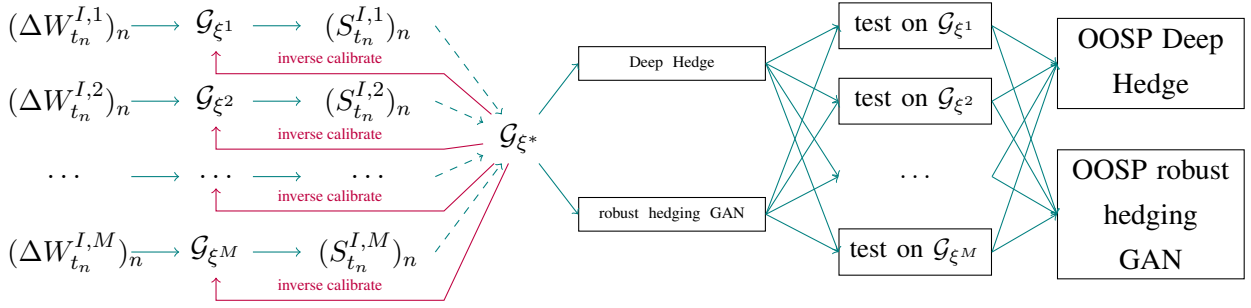
The test structure is visualized in \Cref{fig:TestB}. Usually, the size of $J$ is chosen reasonably high to avoid estimation errors on this side, while the size of $I$ dictates the level of uncertainty the scenarios are generated with. Again, $M$ indicates the precision with which \eqref{id:OOSP2} is evaluated statistically.

\section{Inverse Calibration in the Black-Scholes model}\label{app:InverseCal}

Let $(\bfW_{t_n})_{n \in \myset{0}{N}}$ be a $1$-dimensional Brownian motion under a measure $\bbP$ and denote by $(\Delta \bfW_{t_n})_{n \in \myset{1}{N}}$ its increments, i.e. $\Delta \bfW_{t_n} = \bfW_{t_n} - \bfW_{t_{n-1}}$ for every $n \in \myset{1}{N}$. Then the corresponding Black-Scholes path for parameters $\mu \in \R, \sigma \in \R^+$ is given by
\begin{align*}
    \bfS_n = S_0 \exp\left(\left(\mu - \frac{1}{2}\sigma^2\right) t_n  + \sigma \bfW_n \right), \quad \forall n \in \myset{1}{N},
\end{align*}
what, given the assumption that $\Delta t := t_n - t_{n-1}$  for all $n \in \myset{1}{N}$, implies 
\begin{align}
    \log\left(\frac{\bfS_n}{\bfS_{n-1}} \right) =  \left(\mu - \frac{1}{2}\sigma^2\right) \Delta t  + \sigma \Delta \bfW_n, \quad \forall n \in \myset{1}{N}. \label{id:RelLogReturnsBS}
\end{align}
From this, we see that
\begin{align*}
    \sigma = \sqrt{\frac{1}{\Delta t}\myV{\bbP}{\log\left(\frac{\bfS_n}{\bfS_{n-1}} \right)}}, \quad \forall n \in \myset{1}{N},
\end{align*}
with $\mathbb{V}_\bbP$ being the variance operator under $\bbP$. This derives our estimation formula of $\sigma$, i.e. given a path $(S_n)_{n \in \myset{0}{N}} \in \R^{N + 1}$, we estimate
\begin{align*}
    \hat{\sigma} = \frac{1}{\sqrt{\Delta t}}\widehat{\mathrm{Std}}\left[\left\lbrace\log\left(\frac{S_n}{S_{n-1}} \right)\right\rbrace_{n \in \myset{1}{N}}\right],
\end{align*}
where $\widehat{\mathrm{Std}}[\{\cdot\}_{n \in \myset{1}{N}}]$ is an estimator for the standard deviation. Plugging in \eqref{id:RelLogReturnsBS}, we obtain
\begin{align}
    \hat{\sigma}&= \frac{1}{\sqrt{\Delta t}}\widehat{\mathrm{Std}}\left[\left\lbrace\left(\mu - \frac{1}{2}\sigma^2\right) \Delta t  + \sigma \Delta W_{t_n} \right\rbrace_{n \in \myset{1}{N}}\right], \notag\\
                &= \sigma \widehat{\mathrm{Std}}\left[\left\lbrace Z_{n} \right\rbrace_{n \in \myset{1}{N}}\right], \label{id:InverseCalSigma}
\end{align}
with $(\Delta W_{t_n})_{n \in \myset{1}{N}}$ constituting the sampled Brownian increments that generated the instance of the path $(S_n)_{n \in \myset{0}{N}} \in \R^{N + 1}$ and $Z_n := \Delta W_{t_n} / \sqrt{\Delta t}$ for all $n \in \myset{1}{N}$.

This means, if we want to inverse calibrate $\sigma$ to a fixed $\hat{\sigma}$ for the random noise $(Z_n)_{n \in \myset{1}{N}}$, we obtain
\begin{align*}
    \sigma = \hat{\sigma}\left(\widehat{\mathrm{Std}}\left[\left\lbrace Z_{n} \right\rbrace_{n \in \myset{1}{N}}\right]\right)^{-1}.
\end{align*}

\end{appendices}

\end{document}